\documentclass[reprint,amsmath,amssymb,aip,jcp,eqsecnum]{revtex4-1}

\usepackage{color}

\usepackage{graphicx}      

\newcommand{\beq}{\begin{equation}}
\newcommand{\eeq}{\end{equation}}

\newcommand\beqa{\begin{eqnarray}}
\newcommand\eeqa{\end{eqnarray}}
\newcommand{\nn}{\nonumber\\}

\newcommand{\rr}{\mathbf{r}}

\newcommand{\ex}{\text{ex}}
\newcommand{\hs}{\text{HS}}
\newcommand{\cs}{\text{CS}}
\newcommand{\sw}{\text{SW}}
\newcommand{\shs}{\text{SHS}}
\newcommand{\dl}{\ell}

\def\zero{{(0)}}
\def\one{{(1)}}
\def\two{{(2)}}
\def\three{{(3)}}

\def\n{{(n)}}
\begin{document}

\title{Janus fluid with fixed patch orientations: theory and simulations}

\author{Miguel \'Angel G. Maestre}
\email{maestre@unex.es}
\affiliation{ Departamento de F\'{\i}sica, Universidad de Extremadura,
E-06071 Badajoz, Spain}

\author{Riccardo Fantoni}
\email{rfantoni@ts.infn.it}
\affiliation{Dipartimento di Scienze dei Materiali e Nanosistemi, Universit\`a Ca' Foscari Venezia,
Calle Larga S. Marta DD2137, I-30123 Venezia, Italy}

\author{Achille Giacometti}
\email{achille.giacometti@unive.it}
\affiliation{Dipartimento di Scienze dei Materiali e Nanosistemi, Universit\`a Ca' Foscari Venezia,
Calle Larga S. Marta DD2137, I-30123 Venezia, Italy}

\author{Andr\'es Santos}
\email{andres@unex.es}
\homepage{http://www.unex.es/eweb/fisteor/andres}
\affiliation{ Departamento de F\'{\i}sica, Universidad de Extremadura,
E-06071 Badajoz, Spain}


\date{\today}
\begin{abstract}
We study thermophysical properties of a Janus fluid with constrained orientations, using analytical techniques
and numerical simulations. The Janus character is modeled by means of a Kern--Frenkel potential where
each sphere has one hemisphere of square-well and the other of hard-sphere character. The orientational
constraint is enforced by assuming that each hemisphere can only point either North or South with equal probability.
The analytical approach hinges on a mapping of the above Janus fluid onto
a binary mixture interacting via a ``quasi'' isotropic potential. The anisotropic nature of the original Kern--Frenkel
potential is reflected by the  asymmetry in the interactions occurring between the unlike components of the mixture.
A  rational-function approximation extending the corresponding symmetric case is obtained in the sticky limit, where the
square-well becomes infinitely narrow and deep, and allows a fully analytical approach.
Notwithstanding the rather drastic approximations in the analytical theory, this is shown to provide a rather precise
estimate of the structural and thermodynamical properties of the original Janus fluid.
\end{abstract}
\maketitle
\section{Introduction}
\label{sec:introduction}

Janus fluids refer to colloidal suspensions formed by nearly spherical particles with two different philicities evenly distributed
in the two hemispheres.\cite{Pawar10,Walther08}
Under typical experimental conditions in a water environment, one of the two hemispheres is hydrophobic, while the other { is}
charged, so that different particles tend to repel each other, hence forming isolated monomers. On the other hand, if
repulsive forces are screened
by the addition of a suitable salt, then clusters tend to form driven by hydrophobic interactions.\cite{Hong08}

This self-assembly mechanism has recently attracted increasing attention due to the unprecedented improvement  in the chemical
{ synthesis}
and functionalization of such colloidal particles, that allows a precise and reliable control on the aggregation process
that was not possible until a few years ago.\cite{Bianchi11} From a technological point of view, this is very attractive
as it paves the way to a bottom-up design
and engineering of nanomaterials alternative to conventional top-down techniques.\cite{Zhang04}

One popular choice of model describing the typical duality characteristic of the Janus fluid is the
Kern--Frenkel model.\cite{Kern03}
This model considers a fluid of rigid spheres having their surfaces partitioned into two hemispheres. One of them has a
square-well (SW) character, i.e., it attracts
other \emph{similar} hemispheres through a { SW} interaction, thus mimicking the short-range hydrophobic interactions
occurring in real Janus
fluids. The other part of the surface is assumed to have hard-sphere (HS) interactions with all other hemispheres, i.e.,
with both like HS as well as SW hemispheres. The HS hemisphere hence models the charged part in the limit of highly screened
interactions that
is required to have aggregation of the clusters.

Although in the present paper only an {even distribution between SW and HS surface distributions} will be considered (Janus limit),
other
choices of the coverage, that is the fraction of { SW} surface with respect to the total one, have been studied within the
Kern--Frenkel
model.\cite{Sciortino10} In fact, one of the most attractive { features} of the general model stems from the fact that it
smoothly interpolates between an isotropic HS fluid (zero coverage) and an equally isotropic SW fluid (full coverage).
\cite{Giacometti09,Giacometti10}

The thermophysical and structural properties of the Janus fluid have been recently investigated within the framework of the
Kern--Frenkel model
using numerical simulations,\cite{Sciortino09,Sciortino10} thus rationalizing the cluster formation mechanism characteristic
of the experiments.\cite{Hong08}
The fluid-fluid transition was found to display an unconventional and particularly interesting phase diagram, with a re-entrant
transition associated with
the formation of a cluster phase at low temperatures and densities.\cite{Sciortino09,Sciortino10} While numerical evidence
of this transition is quite convincing,
a minimal theory including all necessary ingredients for {the onset} of this anomalous behavior is still missing. Two previous attempts
are however noteworthy.
Reinhardt \emph{et al.}\cite{Reinhardt11} introduced a van der Waals theory for a suitable mixture of { clusters} and monomers
that accounts for a re-entrant
phase diagram, whereas Fantoni \textit{et al.}\cite{Fantoni11,Fantoni12} developed a cluster theory explaining the appearance of some
``magic numbers'' in the cluster formation.
This notwithstanding, the challenge of an analytical theory fully describing the anomaly { occurring} in the phase diagram of the
Janus fluid still remains.

The aim of the present paper is to { attempt} a new route in this direction. We will do this by considering a Janus fluid within the
Kern--Frenkel model,
where the orientations of the SW hemispheres are constrained to be along either North or South, in a spirit akin to Zwanzig model
for the isotropic-nematic transition in liquid crystals.\cite{Zwanzig63}

Upon observing that under those conditions, one ends up with only four possible different {interactions} (North-North, North-South,
South-North, and South-South),
this constrained model will be further mapped onto a binary mixture interacting via a ``quasi'' isotropic potential. Here the
term ``quasi'' refers to the
fact that a certain memory of the original anisotropic Kern--Frenkel potential is left: after the mapping, one has to discriminate
whether
a particle with patch pointing North (``spin-up'') is lying above or below that with a patch pointing South (``spin-down'').
This will introduce an
{{asymmetry}} in the unlike components of the binary mixture, as explained in detail below.
In order to make the problem tractable from the analytical point of view, the particular limit of an infinitely narrow and deep
square-well (sticky limit) will
be considered. This limit was originally devised by Baxter  and constitutes the celebrated one-component sticky-hard-sphere (SHS)
or adhesive Baxter model.\cite{Baxter68} By construction, our model reduces to it in the limit of fully isotropic attractive
interactions.
The latter model was studied within the Percus--Yevick (PY) closure\cite{Hansen86} in the original Baxter work and in a
subsequent work by Watts
\textit{et al.}\cite{Watts71} The extension of this model to a binary mixture was studied by several
{ authors}.\cite{PS75,Barboy75,Barboy79,Zaccarelli00,TKR02}
{ {The} SHS model with } Kern--Frenkel potential was also studied in Ref.\ \onlinecite{Fantoni07},
{ via a virial expansion at {low densities}}.

A methodology alternative to the one used in the above studies hinges on the so-called ``rational-function approximation''
(RFA),\cite{Santos98,deHaro08} and is known
to be equivalent to the PY approximation for the one-component SHS Baxter model\cite{Baxter68} and for its extension to
symmetric SHS mixtures.\cite{PS75,Santos98,TKR02}
The advantage of this approach { is} that it can be readily extended to more general cases, and this is the reason
why it will be employed in
the present analysis to consider the case of { asymmetric} interactions. We will show that this approach provides a rather
precise estimate {of the thermodynamic and structural properties} of the
Janus fluids with up-down { orientations} by explicitly testing it against Monte Carlo (MC) simulations of the
{ same} Janus fluid.

The remaining part of the paper is envisaged as follows. Section \ref{sec:mapping} describes our Janus model  {{and}} its mapping onto a binary mixture with asymmetric interactions. It is shown in Sec.\ \ref{sec:orientational} that the thermophysical quantities do not require the knowledge of the full (anisotropic) pair correlation functions but only of the functions averaged over all possible {{North or South}} orientations. Section \ref{subsec:sticky} is devoted to the sticky-limit version of the model, i.e., the limit in which the { SW hemisphere has a vanishing well width but an infinite depth leading to} a constant value of the Baxter parameter $\tau$. The exact cavity functions to first order in density (and hence { exact up to} second and third virial coefficients) in the sticky limit are worked out in Appendix \ref{app:appa}. { Up to that} point all the equations are formally exact in the context of the model. Then, in Sec.\ \ref{sec:sticky} we present our approximate RFA theory, which hinges on a heuristic extension from the PY solution for mixtures with symmetric SHS interactions to the realm of asymmetric SHS interactions. Some  technical aspects are relegated to Appendices \ref{appB} and \ref{appC}. The prediction of the resulting analytical theory are compared with MC simulations in Sec.\ \ref{sec:numerical}, where a {{semi-quantitative}} agreement is found. Finally, the paper is closed with conclusions and { an} outlook in Sec.\ \ref{sec:conclusions}.

\section{Mapping the Kern--Frenkel potential onto a binary mixture}
\label{sec:mapping}
\subsection{The Kern--Frenkel potential for a Janus fluid}
\label{subsec:kf}
\begin{figure}
  \includegraphics[width=2.cm]{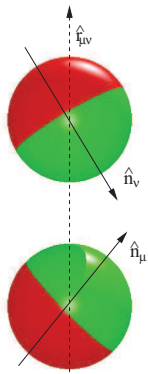}
\caption{The Kern--Frenkel potential for Janus fluids.} \label{fig:fig1}
\end{figure}

{ Consider} a fluid of spheres with identical diameters $\sigma$ where the surface of each sphere is { divided} into two parts.
The first
hemisphere (the green one in the color code given in Fig.\ \ref{fig:fig1}) has a SW character, thus attracting another identical
hemisphere
via a SW potential of width $(\lambda-1) \sigma$ and depth $\epsilon$. The second hemisphere (the red one in the color code of
Fig.\ \ref{fig:fig1})
is instead { a} HS potential.
The orientational dependent pair potential between two arbitrary particles $\mu$ and $\nu$ ($\mu,\nu=1,\ldots,N$, where $N$ is
the total number of particles
in the fluid) has then the form  proposed by Kern and Frenkel\cite{Kern03}
\begin{eqnarray}
\Phi\left(\mathbf{r}_{\mu \nu},\widehat{\mathbf{n}}_{\mu},\widehat{\mathbf{n}}_{\nu}\right) &=& \phi_{\text{HS}} \left(r_{\mu \nu}\right) +\phi_{\text{SW}} \left(r_{\mu \nu}\right) \Psi\left(\widehat{\mathbf{r}}_{\mu \nu},\widehat{\mathbf{n}}_{\mu},
\widehat{\mathbf{n}}_{\nu} \right),\nn
\label{mapping:eq1}
\end{eqnarray}
where the first term is the HS contribution
\begin{equation}
\phi_{\text{HS}}\left(r\right)= \begin{cases}
  \infty,     &   0<r< \sigma ,   \\
0,            &   \sigma < r, \
\end{cases}
 \label{mapping:eq2}
\end{equation}
and the second term is the orientation-dependent attractive part, which can be factorized into an isotropic SW tail
\begin{equation}
\phi_{\text{SW}}\left(r\right)=
\begin{cases}
  - \epsilon, &    \sigma<r< \lambda \sigma ,  \\
0,          &     \lambda \sigma < r,
\end{cases}
 \label{mapping:eq3}
\end{equation}
multiplied by an angular dependent factor

\begin{equation}
\Psi\left(\widehat{\mathbf{r}}_{\mu \nu},\widehat{\mathbf{n}}_{\mu},
\widehat{\mathbf{n}}_{\nu} \right)=
\begin{cases}
  1,    & \text{if }     \widehat{\mathbf{n}}_{\mu} \cdot \widehat{\mathbf{r}}_{\mu \nu} \ge 0  \text{ and }
\widehat{\mathbf{n}}_{\nu} \cdot \widehat{\mathbf{r}}_{\mu \nu} \le 0 \\
0,    &  \text{otherwise} .
\end{cases}
\label{mapping:eq4}
\end{equation}
Here, $\widehat{\mathbf{r}}_{\mu \nu}=\mathbf{r}_{\mu \nu}/r_{\mu \nu}$, where
$\mathbf{r}_{\mu \nu}=\mathbf{r}_{\nu}-\mathbf{r}_{\mu}$, is  the unit vector pointing (by convention) from particle $\mu$ to
particle $\nu$  and the unit vectors $\widehat{\mathbf{n}}_{\mu}$ and $\widehat{\mathbf{n}}_{\nu}$ are ``spin''
vectors associated with the orientation of the attractive hemispheres of particles $\mu$ and $\nu$, respectively
(see Fig.\ \ref{fig:fig1}). An attractive interaction then exists only between the two { SW portions of the surface
sphere,}
provided that the two particles are within the range of the SW potential.

\subsection{{Asymmetric} binary mixture}
\label{subsec:asymmetric}
We now consider the particular case where the only possible orientations of particles are with { attractive caps} pointing only either North
or South with equal probability, as obtained by Fig.\ \ref{fig:fig1} in the limit
$\widehat{\mathbf{n}}_{\mu}=\widehat{\mathbf{z}}$,
$\widehat{\mathbf{n}}_{\nu}=-\widehat{\mathbf{z}}$, { and with $\widehat{\mathbf{z}}$ pointing North}.

\begin{figure}[htbp] 
   \centering
    \includegraphics[width=2.5cm]{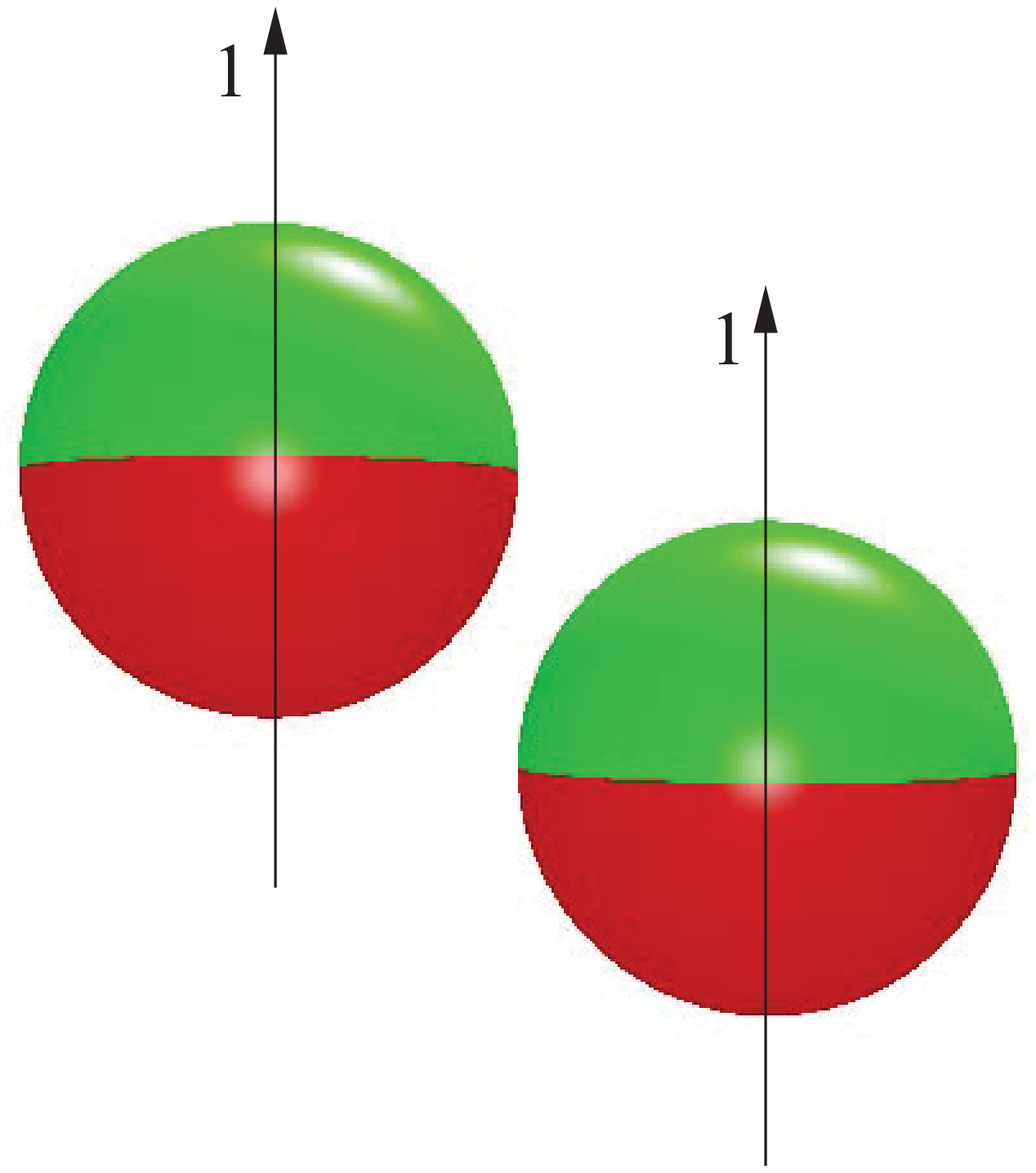}\hfill
   \includegraphics[width=2.5cm]{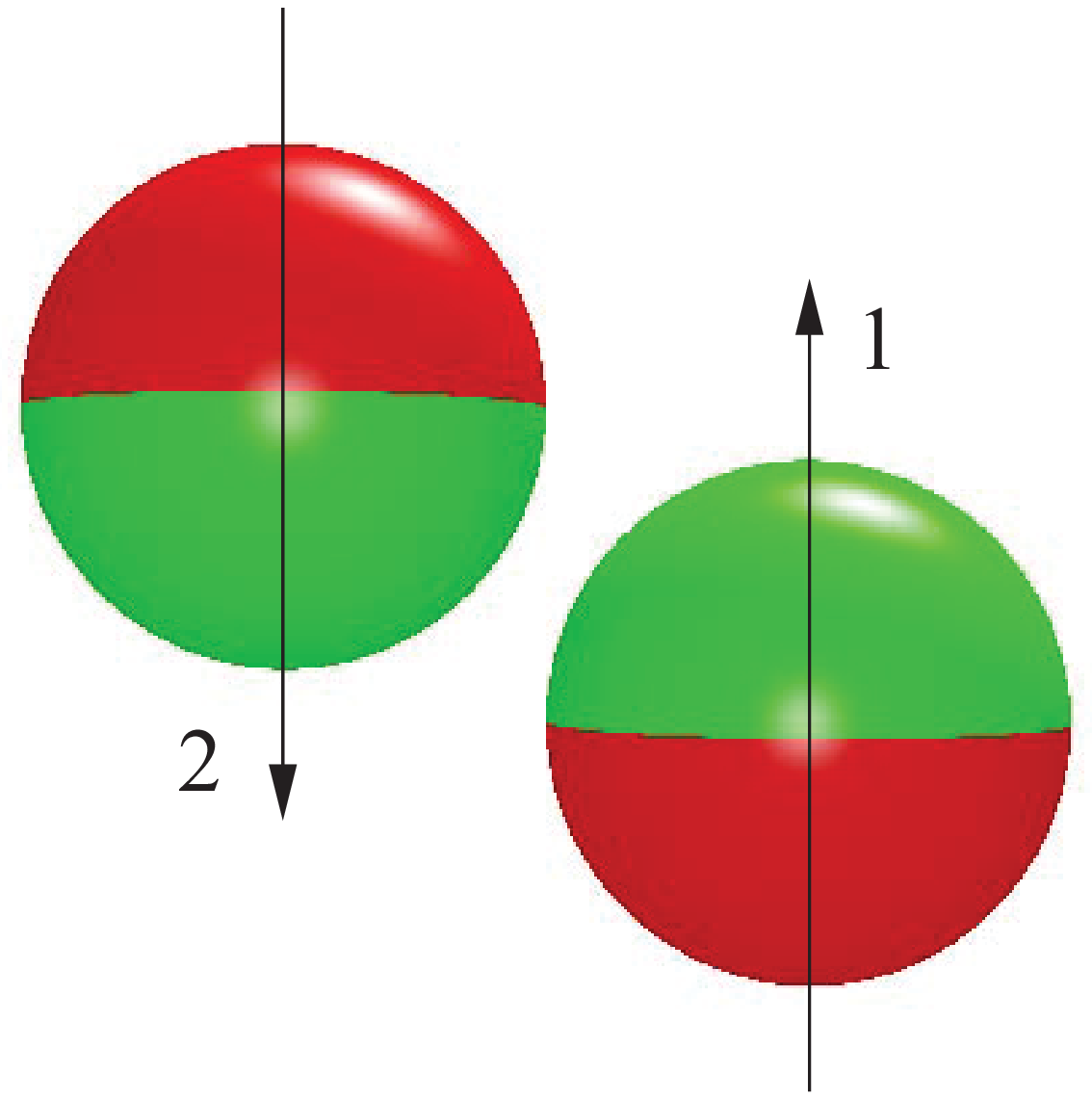}\\
   \vskip1.0cm
      \includegraphics[width=2.5cm]{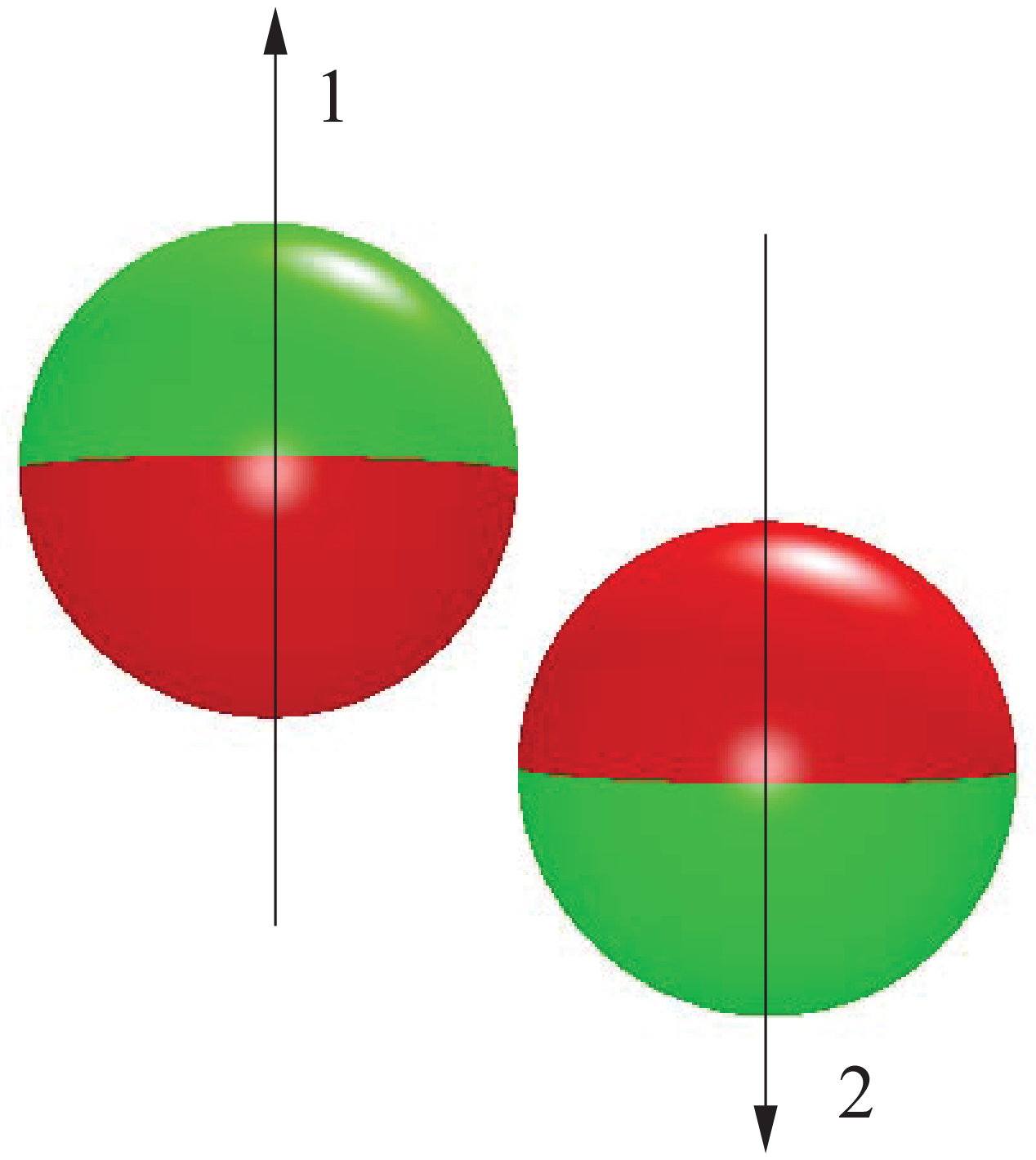}
   \hfill
    \includegraphics[width=2.5cm]{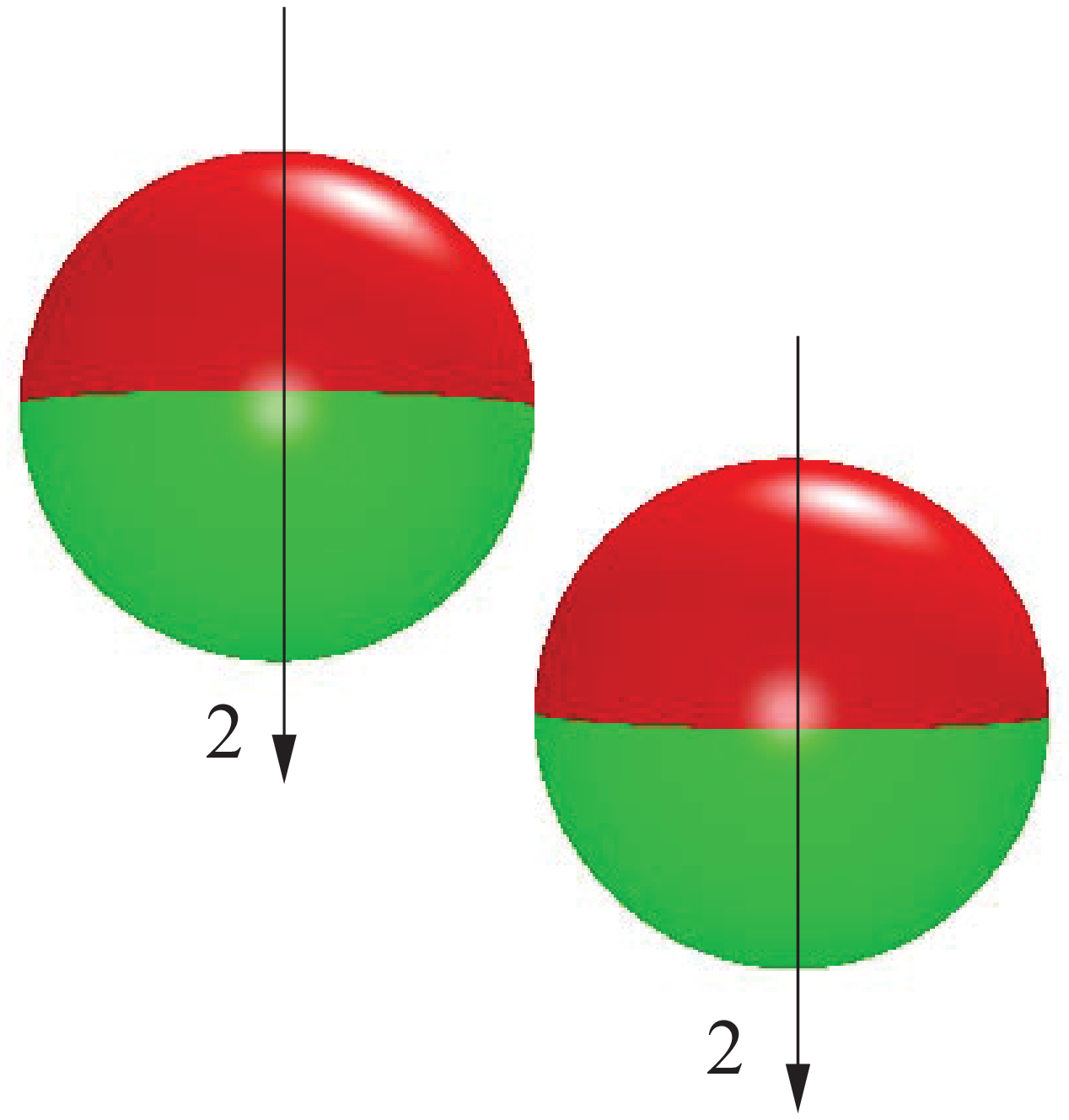}
   \caption{ {
(Top-left) A particle of type $1$ is ``below'' another particle of type $1$ providing SW/HS=HS interactions.
(Top-right) A particle of type $1$ is ``below'' a particle of type $2$ leading to SW/SW=SW interactions.
(Bottom-left) A particle of type $2$ is ``below'' a particle of type $1$ yielding HS/HS=HS interactions.
(Bottom-right) A particle of type $2$ is ``below'' another particle of type $2$ thus leading again to HS/SW=HS interactions.}
} \label{fig:fig2}
\end{figure}

Under these conditions, one then notes that the Kern--Frenkel potential  \eqref{mapping:eq1}--\eqref{mapping:eq4} can be
simplified
by associating a spin $i=1$ (up) to particles with { SW hemispheres} pointing in the North $\widehat{\mathbf{z}}$ direction and a
spin $j=2$ (down) to particles with { SW hemispheres} pointing
in the South $-\widehat{\mathbf{z}}$ direction, so { one is} left with only four possible configurations depending on
whether particles of type $1$ lie
above or below particles of type $2$, as illustrated in Fig.\ \ref{fig:fig2}.
The relationship between the genuine Janus model (see Fig.\ \ref{fig:fig1}) and the { up-down} model (see Fig.\ \ref{fig:fig2}) is reminiscent to the relationship between the { Heisenberg and} the Ising model of ferromagnetism. {}From that point of view, our model can be seen as an Ising-like version of the original Janus model.
{ A similar spirit was also adopted in the Zwanzig model for the isotropic-nematic transition in liquid crystals.\cite{Zwanzig63}}

The advantage of this mapping is that one can { disregard} the original anisotropic Janus-like nature of the interactions and
{recast the problem in
the form of a binary} mixture {such that the interaction potential between a particle of species $i$ located at $\rr_1$ and a particle of species $j$ located at $\rr_2$ has the \emph{asymmetric}  form}
\begin{eqnarray}
\phi_{ij}(\mathbf{r}_{1},\mathbf{r}_{2})&=&\phi_{i j}(\mathbf{r}_{12})\nn
&=&\varphi_{i j}(r_{12})\Theta(\cos\theta_{12})+
\varphi_{j i}(r_{12})\Theta(- \cos\theta_{12}),\nn
\label{mapping:eq5}
\end{eqnarray}
{where $\cos\theta_{12}=\widehat{\mathbf{r}}_{12}\cdot\widehat{\mathbf{z}}$ (recall our convention
$\mathbf{r}_{12}=\mathbf{r}_{2}-\mathbf{r}_{1}$)} and
\beq
\varphi_{ij}(r)=\phi_\hs(r)+\begin{cases}
\phi_\sw(r), & \text{if } i=1 \text{ and } j=2,\\
0,&\text{otherwise}.
\end{cases}
\label{2.1}
\eeq
{ In Eq.\ (\ref{mapping:eq5}) $\Theta(x)=1$ {{and $0$}} for $x>0$ and $x<0$, respectively.}

It is important to remark that, in general, $\varphi_{12}(r)\neq \varphi_{21}(r)$, { as evident from Eq.\ (\ref{2.1})}. Thus, the binary
mixture is not necessarily symmetric [unless $\epsilon=0$ or $\lambda=1$ in Eq.\ \eqref{mapping:eq3}],
unlike standard binary mixtures where this symmetry condition is ensured by construction.
In the  potential \eqref{mapping:eq5}, { there however is} still a ``memory''
of the original anisotropy since
the potential energy of a pair of particles of species $i$ and $j$ separated a distance {$r_{12}$} depends on whether particle $j$
is ``above'' {($\cos\theta_{12}>0$)} or ``below'' {($\cos\theta_{12}<0$)} particle $i$.
In this sense, the binary mixture obtained in this way
is ``quasi'', and not ``fully'', spherically symmetric.

\begin{figure}[htbp]
\centering
\includegraphics[width=8.5cm]{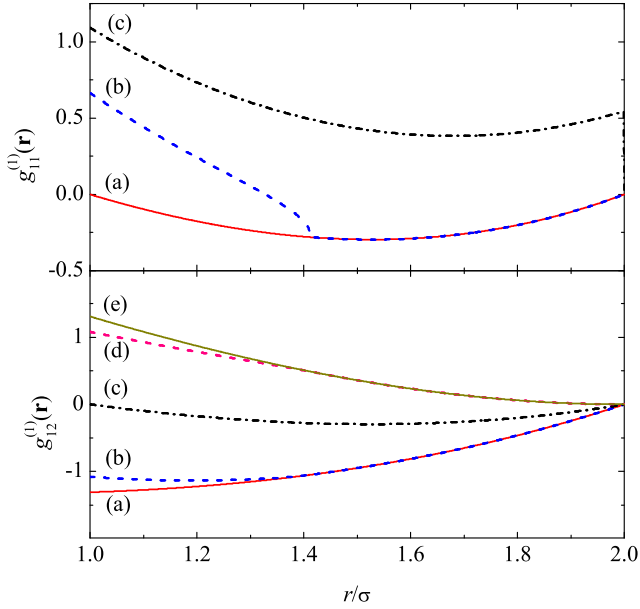}
\caption{{(Top) Plot of $g_{11}^\one(\mathbf{r})$ as a function of $r$ for (a) $\theta=0$ and $\pi$, (b) $\theta=\frac{\pi}{4}$ and $\frac{3\pi}{4}$, and (c) $\theta=\frac{\pi}{2}$. (Bottom) Plot of the regular part of $g_{12}^\one(\mathbf{r})$ as a function of $r$ for (a) $\theta=0$, (b) $\theta=\frac{\pi}{4}$,  (c) $\theta=\frac{\pi}{2}$, (d) $\theta=\frac{3\pi}{4}$, and (e) $\theta={\pi}$. The interaction potential is given by Eq.\ \eqref{2.1} (sketched in Fig.\ \protect\ref{fig:fig2}), except that the sticky limit with Baxter's temperature $\tau=0.1$ has been taken (see Sec.\ \ref{subsec:sticky}).}}
\label{fig:fig_y}
\end{figure}

Another important point to be stressed is that, while  the \emph{sign} of
$\cos \theta_{12}$ represents the only source of
anisotropy in the above potential $\phi_{i j}(\mathbf{r}_{12})$, this is \emph{not} the case for the corresponding correlation
functions, which will explicitly depend upon the relative
orientation $\cos \theta_{12}$
and not only upon its sign. This applies, for instance, to the pair correlation functions $g_{ij}(\mathbf{r})=g_{ij}(r;\theta)$, as shown
in Appendix \ref{app:appa} to first order in density in the sticky limit (see Sec.\ \ref{subsec:sticky}). As an illustration, Fig.\ \ref{fig:fig_y} shows the first-order pair correlation functions $g_{11}^\one(\mathbf{r})$ and $g_{12}^\one(\mathbf{r})$ as functions of the radial distance $r$ for several orientations $\theta$.

As our aim is to remove the orientational dependence in the original potential altogether, a further
simplification is required to reduce the problem
to a simple binary mixture having \emph{asymmetric} correlation functions dependent only on distances and not on
orientations of spheres.
This will be the orientational average discussed in Sec.\ \ref{sec:orientational}.

\section{Orientational average and thermodynamics}
\label{sec:orientational}
\subsection{Orientational average}
\label{subsec:orientational}

Most of the content of this section applies to a mixture (with any number of components) characterized by any anisotropic potential $\phi_{ij}(\mathbf{r})=\phi_{ji}(-\mathbf{r})$
exhibiting the quasi-isotropic form (\ref{mapping:eq5}), where, in general $\varphi_{ij}(r)\neq \varphi_{ji}(r)$ if $i\neq j$.
In that case, we note that the thermodynamic quantities will generally involve integrals of the general form
\begin{eqnarray}
I_{ij}&=&\int d\mathbf{r}\, g_{ij}(\mathbf{r})\mathcal{F}_{ij}(\mathbf{r})
\label{orientational:eq1}
\end{eqnarray}
with
\begin{eqnarray}
\mathcal{F}_{ij}\left(\mathbf{r}\right)&=&{F}_{ij}(r)\Theta\left(\cos\theta\right)+{F}_{ji}(r)\Theta\left(-\cos\theta\right),
\label{orientational:eq2}
\end{eqnarray}
where in general $F_{ij}(r)\neq F_{ji}(r)$ if $i\neq j$.
This strongly suggests { that} one can define the two orientational \emph{averages} $g_{ij}^+(r)$ and $g_{ij}^-(r)$ as
\beq
g_{ij}^+(r)\equiv\overline{g}_{ij}(r)={\int_{0}^{1} d \left(\cos \theta \right)
 g_{ij}(\mathbf{r})},
\label{orientational:eq3a}
\eeq
\beq
 g_{ij}^-(r)\equiv\overline{g}_{ji}(r)={\int_{-1}^{0} d \left(\cos \theta \right)
 g_{ij}(\mathbf{r})}
.
\label{orientational:eq3b}
\eeq
Note that $g_{ij}^-(r)=g_{ji}^+(r)${, and this suggests the use of} the notation $\overline{g}_{ij}(r)$ and
$\overline{g}_{ji}(r)$ instead of $g_{ij}^+(r)$ and $g_{ij}^-(r)$, respectively.
Taking into account Eqs.\ \eqref{orientational:eq2}--\eqref{orientational:eq3b}, { Eq.}\
\eqref{orientational:eq1} becomes
\begin{eqnarray}
I_{ij}&=&\frac{1}{2}\int d\mathbf{r}\,\left[\overline{g}_{ij}(r){F}_{ij}\left(r\right)+
\overline{g}_{ji}(r){F}_{ji}\left(r\right)\right].
\label{orientational:eq4}
\end{eqnarray}
In the case of a double summation over $i$ and $j$,
\begin{eqnarray}
\sum_{i,j} x_i x_j I_{ij}&=&\sum_{i,j} x_i x_j \int d\mathbf{r}\,\overline{g}_{ij}(r){F}_{ij}\left(r\right),
\label{orientational:eq5}
\end{eqnarray}
where $x_i$ denotes the mole fraction of species $i$.

\subsection{Thermodynamics of the mixture: energy, virial, and compressibility routes}
\label{subsec:thermodynamics}

We can now particularize the { general} result (\ref{orientational:eq5}) to specific cases.

In the case of the internal { energy},  $\mathcal{F}_{ij}(\mathbf{r})=\phi_{ij}(\mathbf{r})$ and so the energy equation of
state can be written as\cite{Hansen86}
\begin{eqnarray}
\label{orientational:eq6}
u_{\text{ex}}&=&
\frac{1}{2}\rho\sum_{i,j}
x_i x_j \int d{\bf r}\,  g_{ij}\left(\mathbf{r}\right)\phi_{ij}\left(\mathbf{r}\right)\nn
&=&
\frac{1}{2}\rho\sum_{i,j}
x_i x_j \int d{\bf r}\,  \overline{y}_{ij}\left(r\right)\varphi_{ij}(r)e^{-\beta\varphi_{ij}(r)},
\end{eqnarray}
where $u_{\text{ex}}$ is the excess internal energy per particle, $\rho$ is the number density, $\beta=1/k_BT$ ($k_B$ and $T$ being the Boltzmann constant
and the temperature, respectively), and
$\overline{y}_{ij}({r})\equiv \overline{g}_{ij}({r})e^{\beta\varphi_{ij}({r})}$ is the orientational average of the cavity
function
${y}_{ij}(\mathbf{r})\equiv {g}_{ij}(\mathbf{r})e^{\beta\phi_{ij}(\mathbf{r})}$.

A similar result holds for the virial route to the equation of state,
\begin{eqnarray}
\label{orientational:eq7}
Z&\equiv& \frac{P}{\rho k_BT}\nn
&=&1+\frac{1}{6}\rho\sum_{i,j}
x_i x_j \int d\mathbf{r}\, y_{ij}\left(\mathbf{r}\right)\mathbf{r}\cdot\nabla e^{-\beta\phi_{ij}\left(\mathbf{r}\right)},
\end{eqnarray}
where $P$ is the pressure.
First, note that
\begin{eqnarray}
\label{orientational:eq8}
\nabla \phi_{ij}\left(\mathbf{r}\right)&=&\left[\frac{d\varphi_{ij}\left(r\right)}{dr}\Theta\left(\cos\theta\right)
+\frac{d\varphi_{ji}\left(r\right)}{dr}\Theta\left(-\cos\theta\right)\right]\widehat{\mathbf{r}}\nn
&&-\delta\left(z\right)\left[\varphi_{ij}\left(r\right)-\varphi_{ji}\left(r\right)\right]\widehat{\mathbf{z}}.
\end{eqnarray}
Therefore,
\begin{eqnarray}
\label{orientational:eq9}
\mathbf{r}\cdot\nabla \phi_{ij}(\mathbf{r})=r \left[\frac{d\varphi_{ij}(r)}{dr}\Theta( \cos\theta)
+\frac{d\varphi_{ji}(r)}{dr}\Theta(-\cos\theta)\right],\nn
\end{eqnarray}
and thus
\begin{eqnarray}
\label{orientational:eq10}
Z=1+\frac{1}{6}\rho\sum_{i,j}
x_i x_j \int d\mathbf{r}\, \overline{y}_{ij}\left({r}\right){r}\frac{d}{dr} e^{-\beta\varphi_{ij}({r})}.
\end{eqnarray}

Finally, let us consider the compressibility route. In a mixture, the (dimensionless) isothermal compressibility $\chi_T$  is in general given by
\begin{eqnarray}
\label{orientational:eq11}
\chi_{T}^{-1}&=&\frac{1}{k_BT}\left(
\frac{\partial p}{\partial\rho}\right)_{T,\{x_j\}}\nn
&=&
\sum_{i,j}\sqrt{x_ix_j}
\left[\mathsf{I}+\widehat{\mathsf{h}}\left(0\right)\right]^{-1}_{ij},
\end{eqnarray}
where $\widehat{h}_{ij}(0)$ is proportional to
the zero wavenumber limit of the Fourier transform of the total correlation function $h_{ij}(\mathbf{r})=g_{ij}(\mathbf{r})-1$, namely
\begin{eqnarray}
\widehat{h}_{ij}(0)&=&\rho\sqrt{x_i x_j}\int d\mathbf{r}\,h_{ij}\left(\mathbf{r}\right)\nn
&=&\frac{\rho\sqrt{x_i x_j}}{2}\int d\mathbf{r}\,\left[\overline{h}_{ij}\left({r}\right)+\overline{h}_{ji}\left({r}\right)\right].
\label{orientational:eq12}
\end{eqnarray}
In the specific case of a binary mixture considered here, Eq.\ \eqref{orientational:eq11} becomes
\beq
\label{orientational:eq13}
\chi_{T}^{-1}=\frac
{1+x_2\widehat{h}_{11}(0)+ x_1\widehat{h}_{22}(0)-2 \sqrt{x_1 x_2}\widehat{h}_{12}(0)}{\left[1+ \widehat{h}_{11}(0)\right]
\left[1+ \widehat{h}_{22}(0)\right]
- \widehat{h}_{12}^2(0)}.
\eeq

Equations \eqref{orientational:eq6}, \eqref{orientational:eq10}, \eqref{orientational:eq11}, and \eqref{orientational:eq12}
confirm that the knowledge of the two average quantities $\overline{g}_{ij}(r)$ and $\overline{g}_{ji}(r)$ for each pair $ij$ suffices to
determine the thermodynamic quantities. In fact, Eqs.\ \eqref{orientational:eq6}, \eqref{orientational:eq10},
\eqref{orientational:eq11}, and \eqref{orientational:eq12} are formally indistinguishable from those corresponding to mixtures with
standard isotropic interactions, except that in our case one generally has $\varphi_{ij}(r)\neq \varphi_{ji}(r)$ and, consequently,
$\overline{g}_{ij}(r)\neq \overline{g}_{ji}(r)$.

For future convenience, it is expedient to introduce the Laplace transform of $r\overline{g}_{ij}(r)$:
\begin{eqnarray}
\label{orientational:eq14}
G_{ij}(s)&=&\int_0^\infty dr\, e^{-sr}r \overline{g}_{ij}(r).
\end{eqnarray}
Its small-$s$ behavior is of the form\cite{deHaro08}
\begin{eqnarray}
\label{orientational:eq15}
s^2 G_{ij}(s)&=&1+H_{ij}^{(0)}s^2+H_{ij}^{(1)}s^3+\cdots,
\end{eqnarray}
where
\begin{eqnarray}
\label{orientational:eq16}
H_{ij}^\n\equiv \int_0^\infty dr\, (-r)^n r \overline{h}_{ij}(r).
\end{eqnarray}
Thus, Eq.\ \eqref{orientational:eq12} becomes
\begin{eqnarray}
\widehat{h}_{ij}(0)=-2\pi\rho\sqrt{x_i x_j}\left[H_{ij}^\one+H_{ji}^\one\right].
\label{orientational:eq17}
\end{eqnarray}

Note that Eq.\ \eqref{orientational:eq15} implies
\beq
\lim_{s\to 0}s^2 G_{ij}(s)=1,
\label{III.1a}
\eeq
\beq
\lim_{s\to 0}\frac{s^2 G_{ij}(s)-1}{s}=0.
\label{III.1b}
\eeq
\section{The sticky limit}
\label{subsec:sticky}
The mapping of the { Kern--Frenkel} potential with fixed patch orientation along the $\pm \widehat{\mathbf{z}}$ axis onto
a binary mixture represents a considerable simplification. On the other hand{,} no approximation is involved
in { this} mapping.

The presence of the original SW interactions for the radial part [see Eq.\ \eqref{mapping:eq3}] makes the analytical treatment of the problem  a formidable task.
Progresses can however be made by considering the Baxter { SHS} limit, for which a well defined approximate scheme of solution
is available in the \emph{isotropic} case for both one-component\cite{Baxter68} and multi-component { \cite{PS75,Barboy75,Barboy79,Zaccarelli00,TKR02}} fluids. The discussion reported below closely follows the analogue for Baxter { symmetric} mixtures.\cite{Barboy75,Barboy79}

Let us start by rewriting Eq.\ \eqref{2.1} as
\begin{eqnarray}
\label{sticky:eq1}
\varphi_{ij}(r)=\left\{
\begin{array}{ll}
\infty,&r<\sigma,\\
-\epsilon_{ij},&\sigma<r<\lambda\sigma,\\
0,&r>\lambda\sigma,
\end{array}
\right.
\end{eqnarray}
where $\epsilon_{11}=\epsilon_{22}=\epsilon_{21}=0$ and $\epsilon_{12}=\epsilon>0$. However, in this section we will assume generic energy scales $\epsilon_{ij}$. In that case, the virial equation of state \eqref{orientational:eq10} becomes
\beq
\label{analytical:eq1}
Z=1+4\eta
\overline{y}(\sigma)-12\eta\sum_{i,j} x_i x_j{t_{ij}}
\frac{\lambda^3 \overline{y}_{ij}(\lambda\sigma)-
\overline{y}_{ij}(\sigma)}{\lambda^3-1},
\eeq
where $\eta\equiv \frac{\pi}{6}\rho\sigma^3$ is the packing fraction,
\beq
\overline{y}(r)=\sum_{i,j}x_i x_j
\overline{y}_{ij}(r)
\label{global}
\eeq
is the orientational average \emph{global} cavity function,
 and
\begin{eqnarray}
\label{analytical:eq2}
t_{ij}\equiv\frac{1}{12\tau_{ij}}&\equiv&
\frac{1}{3}\left(\lambda^3-1\right)
\left(e^{\beta\epsilon_{ij}}-1\right)
\end{eqnarray}
is a parameter measuring the degree of ``stickiness'' of the SW interaction
$\varphi_{ij}(r)$. This parameter will be used later on to connect results from numerical simulations of the { actual} Janus fluid with analytical results derived for asymmetric SHS mixtures. Although Baxter's temperature parameters $\tau_{ij}$ are commonly used in the literature, we will employ the inverse temperature parameters $t_{ij}=1/12\tau_{ij}$ in most of the mathematical expressions.

In the case of the interaction potential \eqref{sticky:eq1}, the energy equation of state \eqref{orientational:eq6} reduces to
\begin{eqnarray}
\label{analytical:eq3}
u_{\text{ex}}&=&-12\frac{\eta}{\sigma^3}\sum_{i,j} x_i x_j \epsilon_{ij}
e^{\beta\epsilon_{ij}}
\int_{\sigma}^{\lambda\sigma} dr\,r^2\overline{y}_{ij}(r).
\end{eqnarray}
The compressibility equation of state \eqref{orientational:eq11} does not simplify for the SW
interaction {.}

Since the (orientational average) cavity function $\overline{y}_{ij}(r)$ must be continuous, it follows that
\beq
\label{analytical:eq4}
\overline{g}_{ij}(r)=\overline{y}_{ij}(r)\left[e^{\beta\epsilon_{ij}}\Theta(r-\sigma)
-\left(e^{\beta\epsilon_{ij}}-1\right)\Theta(r-\lambda\sigma)\right].
\eeq

Following Baxter's prescription,\cite{Baxter68} we now consider the SHS limit
\beq
\label{analytical:eq5}
\lambda\to 1, \quad \epsilon_{ij}\to\infty,\quad t_{ij}\equiv \frac{1}{12\tau_{ij}}\to
{(\lambda-1)}{e^{\beta\epsilon_{ij}}}
=\text{finite},
\eeq
so that the well \eqref{sticky:eq1} becomes infinitely deep and narrow and can be described by a single (inverse) stickiness parameter $\tau_{ij}$.
Note that in the { present} Janus case ($\epsilon_{11}=\epsilon_{22}=\epsilon_{21}=0$, $\epsilon_{12}=\epsilon>0$) one actually has $t_{11}=t_{22}=t_{21}=0$ and $t_{12}=t=1/12\tau$.

In the SHS limit \eqref{analytical:eq5}, Eqs.\ (\ref{analytical:eq1}), \eqref{analytical:eq3}, and (\ref{analytical:eq4}) become, respectively,
\beq
\label{analytical:eq6}
Z=1+4\eta
\overline{y}(\sigma)-4\eta\sum_{i,j} x_i x_j
{t_{ij}}\left[3\overline{y}_{ij}(\sigma)+\sigma \overline{y}_{ij}'(\sigma)\right]
,
\eeq
\begin{eqnarray}
\label{analytical:eq7}
u_{\text{ex}}&=&-12{\eta}\sum_{i,j} x_i x_j
\epsilon_{ij}t_{ij}\overline{y}_{ij}(\sigma),
\end{eqnarray}
\begin{eqnarray}
\label{analytical:eq8}
\overline{g}_{ij}(r)&=&\overline{y}_{ij}(r)\left[\Theta(r-\sigma)
+{t_{ij}}\sigma\delta_+(r-\sigma)\right].
\end{eqnarray}
In Eq.\ \eqref{analytical:eq6},
$\overline{y}_{ij}'(\sigma)$ must be interpreted as $\lim_{\lambda\to 1}\left.\frac{d}{dr}\overline{y}_{ij}(r)\right|_{r=\sigma}$, which in principle differs from
$\left.\frac{d}{dr}\lim_{\lambda\to 1}\overline{y}_{ij}(r)\right|_{r=\sigma}$.\cite{YS93}
However, both quantities coincide in the one-dimensional case\cite{YS93} and are expected to coincide in the three-dimensional case as well. This is just a consequence of the expected continuity of $\frac{d}{dr}\overline{y}_{ij}(r)$ at $r=\lambda\sigma$ in the SW case.\cite{A00}

Thermodynamic consistency between the virial and energy routes implies
\beq
\rho\frac{\partial u_\ex}{\partial\rho}=\frac{\partial Z}{\partial\beta}=\sum_{i,j}\epsilon_{ij}t_{ij}\frac{\partial Z}{\partial t_{ij}}.
\label{analytical:eq42}
\eeq
Using Eqs.\ \eqref{analytical:eq6} and \eqref{analytical:eq7} and equating the coefficients of $\epsilon_{ij}$ in both sides, { the consistency
condition (\ref{analytical:eq42}) yields}
\beqa
&&
x_i x_j\left[\sigma \overline{y}_{ij}'(\sigma)-3\eta\frac{\partial \overline{y}_{ij}(\sigma)}{\partial\eta}\right]=\sum_{k,\ell}x_kx_\ell\nn
&&
\times\left\{\frac{\partial\overline{y}_{k\ell}(\sigma)}{\partial t_{ij}}-
{t_{k\ell}}\frac{\partial}{\partial t_{ij}}\left[3\overline{y}_{k\ell}(\sigma)+\sigma \overline{y}_{k\ell}'(\sigma)\right]
\right\}.
\label{consistency}
\eeqa

For distances $r\gtrsim \sigma$, the orientational averages of the cavity functions can be { Taylor expanded as}
\beq
\label{analytical:eq11}
\Theta(r-\sigma) \overline{y}_{ij}(r)=\Theta(r-\sigma)\left[\overline{y}_{ij}(\sigma)+\overline{y}_{ij}'(\sigma)(r-\sigma)+\cdots\right].
\eeq
{ Hence}, if we denote by $Y_{ij}(s)$ the Laplace transform of $\Theta(r-\sigma)r \overline{y}_{ij}(r)$, Eq.\ \eqref{analytical:eq11}
{ yields for large $s$}
\beq
\label{analytical:eq12}
e^{\sigma s} s Y_{ij}(s)=\sigma\overline{y}_{ij}(\sigma)+\left[\overline{y}_{ij}(\sigma)+\sigma\overline{y}_{ij}'(\sigma)\right]s^{-1}+\cdots.
\eeq
According to  { Eqs.\ \eqref{analytical:eq8} and \eqref{orientational:eq14}}, the relationship between the Laplace function $G_{ij}(s)$ and $Y_{ij}(s)$ is
\begin{eqnarray}
\label{analytical:eq10}
G_{ij}(s)&=&Y_{ij}(s)+{\sigma^2}t_{ij}\overline{y}_{ij}(\sigma)e^{-\sigma s}.
\end{eqnarray}
Inserting Eq.\ \eqref{analytical:eq12} into Eq.\ \eqref{analytical:eq10}, we obtain the following
large-$s$ behavior of $G_{ij}(s)$:
\begin{eqnarray}
\label{analytical:eq13}
e^{\sigma s}G_{ij}(s)&=&
{\sigma^2}t_{ij}\overline{y}_{ij}(\sigma)+\sigma \overline{y}_{ij}(\sigma)s^{-1}\nn
&&+\left[\overline{y}_{ij}(\sigma)+\sigma\overline{y}_{ij}'(\sigma)\right]s^{-2}
+{\cal O}(s^{-3}).\nn
\end{eqnarray}
A consequence of this is
\beq
\frac{\lim_{s\to\infty}e^{\sigma s}G_{ij}(s)}{\lim_{s\to\infty}s\left[e^{\sigma s}G_{ij}(s)-\lim_{s\to\infty}e^{\sigma s}G_{ij}(s)\right]}=\sigma t_{ij}.
\label{IV.1}
\eeq

\section{A heuristic, non-perturbative analytical theory}
\label{sec:sticky}
\subsection{A simple approximate scheme within the Percus--Yevick closure}
\label{subsec:simple}
The Ornstein--Zernike (OZ) equation for an anisotropic mixture reads\cite{Hansen86}
\begin{eqnarray}
\label{analytical:eq14}
{h_{ij}(\mathbf{r}_{12})}&=&{c_{ij}(\mathbf{r}_{12})+\rho\sum_{k} x_k \int d\mathbf{r}_3\, h_{ik}(\mathbf{r}_{13})c_{kj}(\mathbf{r}_{32})}\nn
&=&{c_{ij}(\mathbf{r}_{12})+\rho\sum_{k} x_k \int d\mathbf{r}_3\, c_{ik}(\mathbf{r}_{13})h_{kj}(\mathbf{r}_{32})},\nn
\end{eqnarray}
where {$c_{ij}(\mathbf{r})$} is the direct correlation function. The { PY closure reads}
\begin{eqnarray}
\label{analytical:eq15}
c_{ij}(\mathbf{r})&=&g_{ij}(\mathbf{r})\left[1-e^{\beta\phi_{ij}(\mathbf{r})}\right].
\end{eqnarray}
{ Introducing the averages $c_{ij}^+(r)=\overline{c}_{ij}(r)$ and $c_{ij}^-(r)=\overline{c}_{ji}(r)$ for} $c_{ij}(\mathbf{r})$ in a way similar to Eqs.\ \eqref{orientational:eq3a} and \eqref{orientational:eq3b},
Eq.\ \eqref{analytical:eq15} yields
\begin{eqnarray}
\label{analytical:eq16}
\overline{c}_{ij}({r})=\overline{g}_{ij}({r})\left[1-e^{\beta\varphi_{ij}({r})}\right].
\end{eqnarray}
Thus, the PY closure for the full correlation functions $c_{ij}(\mathbf{r})$ and $g_{ij}(\mathbf{r})$ translates into an equivalent relation for the orientational average functions $\overline{c}_{ij}({r})$ and $\overline{g}_{ij}({r})$.
A similar reasoning, on the other hand, is not valid for the OZ relation. Multiplying both sides of the first equality in Eq.\  \eqref{analytical:eq14} by {$\Theta(\cos\theta_{12})$ }and integrating over {$\cos\theta_{12}$} one gets
\beqa
\label{analytical:eq17}
{\overline{h}_{ij}({r}_{12})}&=&{\overline{c}_{ij}({r}_{12})+\rho\sum_{k} x_k \int d\mathbf{r}_3\, \int_0^{1} d\left(\cos\theta_{12}\right)}\nn
&&\times
{h_{ik}(\mathbf{r}_{13})c_{kj}(\mathbf{r}_{32})}.
\eeqa
The same result is obtained if we start from the second equality in Eq.\  \eqref{analytical:eq14}, multiply by {$\Theta(-\cos\theta_{12})$}, integrate over {$\cos\theta_{12}$}, and make the changes {$\mathbf{r}_{12}\to -\mathbf{r}_{12}$, $\mathbf{r}_{13}\to -\mathbf{r}_{13}$}, and $i\leftrightarrow j$.
Equation \eqref{analytical:eq17} shows that in the case of anisotropic potentials of the form \eqref{mapping:eq5} the OZ equation does not reduce to a closed equation involving the averages {$\overline{h}_{ij}({r})$} and {$\overline{c}_{ij}({r})$} only{, as remarked.

 In order to obtain {{a}} closed theory,  we adopt} the \emph{heuristic} mean-field decoupling approximation
\begin{eqnarray}
\label{analytical:eq17b}
&&{\int d\mathbf{r}_3\, \int_0^{1} d\left(\cos \theta_{12}\right)\,h_{ik}(\mathbf{r}_{13})c_{kj}(\mathbf{r}_{32})}\nn
\to&&
{\int d\mathbf{r}_3\,\overline{h}_{ik}\left({r}_{13}\right)}{\overline{c}_{kj}\left({r}_{32}\right)}.
\end{eqnarray}
Under these conditions, the true OZ relation \eqref{analytical:eq17} is replaced by the pseudo-OZ relation
\beq
\label{analytical:eq18}
{\overline{h}_{ij}({r}_{12})=\overline{c}_{ij}({r}_{12})+\rho\sum_{k} x_k \int d\mathbf{r}_3\, \overline{h}_{ik}({r}_{13})\overline{c}_{kj}({r}_{32})}.
\eeq
This can { then be closed by the PY equation \eqref{analytical:eq16} and standard theory applies. An alternative and equivalent view is to consider $\overline{c}_{ij}({r})$ not as the orientational average of the true direct correlation function {$c_{ij}(\mathbf{r})$} but as exactly defined by Eq.\ \eqref{analytical:eq18}. Within this interpretation, Eq.\ \eqref{analytical:eq16} then represents} a pseudo-PY closure not derivable from the true PY closure \eqref{analytical:eq15}.

{ Within the above interpretation, it} is then important to bear in mind that the functions $\overline{g}_{ij}(r)$ obtained from the solution of { a combination of} Eqs.\ \eqref{analytical:eq16} and
\eqref{analytical:eq18} are \emph{not} equivalent to the orientational averages of the functions $g_{ij}(\mathbf{r})$ obtained from the solution of the true PY problem
posed by Eqs.\ \eqref{analytical:eq14} { and} complemented by the PY condition \eqref{analytical:eq15}.
As a consequence, the solutions to Eqs.\ \eqref{analytical:eq16} and
\eqref{analytical:eq18} are \emph{not} expected to provide the exact $\overline{g}_{ij}(r)$ to first order in $\rho$, in contrast to the true PY problem.
{ This is an interesting nuance that will be further discussed in Sec.\ \ref{subsec_LDE}.}

The main advantage of the approximate OZ relation \eqref{analytical:eq18} in the case of anisotropic interactions of the form \eqref{mapping:eq5}
is that it allows to transform the obtention of an \emph{anisotropic} function $g_{ij}(\mathbf{r})$, but \emph{symmetric} in the sense {that}
$g_{ij}(\mathbf{r})=g_{ji}(-\mathbf{r})$, into the obtention of an \emph{isotropic} function $\overline{g}_{ij}(r)$, but \emph{asymmetric}
since $\overline{g}_{ij}(r)\neq \overline{g}_{ji}(r)$. In the case of the anisotropic SHS potential defined above, we can exploit the known solution of
the PY equation for \emph{isotropic} SHS mixtures to construct the solution of the set made of Eqs.\ \eqref{analytical:eq16} and \eqref{analytical:eq18}.
This is done in  Subsection \ref{subsec:rfa} by following the RFA methodology.

\subsection{RFA method for SHS}
\label{subsec:rfa}
Henceforth, { for the sake of simplicity,} we take $\sigma=1$ as length unit.
The aim of this section is to extend the RFA approximation proposed for symmetric SHS mixtures\cite{Santos98,deHaro08} to the { asymmetric} case.

We start with the one-component case.\cite{YS93} Let us introduce an auxiliary function $F(s)$ related to the Laplace transform $G(s)$ of $rg(r)$ by
\beq
G(s)=\frac{1}{2\pi}\frac{se^{-s}}{F(s)+\rho e^{-s}}.
\label{F(s)}
\eeq
{The next} step is to approximate $F(s)$ by a \emph{{rational function}},
\beq
F(s)=\frac{S(s)}{L(s)},
\eeq
with  $S(s)=S^\zero+S^\one s+S^\two s^2+s^3$ and
\beq
L(s)=L^\zero+L^\one s+L^\two s^2.
 \label{L(s)}
\eeq
Note that $\lim_{s\to\infty} F(s)/s=1/L^\two=\text{finite}$, so that $\lim_{s\to\infty} e^s G(s)=\text{finite}$, in agreement with Eq.\ \eqref{analytical:eq13}. Furthermore, Eq.\ \eqref{III.1a}  { requires}  $F(s)+\rho e^{-s}=\mathcal{O}(s^3)$, so that
$S^\zero=-\rho L^\zero$, $S^\one=\rho\left(L^\zero-L^\one\right)$, $S^\two=\rho\left(L^\one-\frac{1}{2}L^\zero-L^\two\right)$.
Taking all of this into account, Eq.\ \eqref{F(s)} can be rewritten as
\beq
G(s)=\frac{e^{- s}}{2\pi s^2}\frac{{L}(s)}{1-A(s)},
\label{Gpure}
\eeq
where
\beq
A(s)=\frac{\rho}{s^3}\left[(1-e^{-s})L(s)-L^\zero s+\left(\frac{1}{2}L^\zero-L^\one\right)s^2\right].
\label{A(s)}
\eeq

{ In} the case of a mixture, $G(s)$, $L(s)$, and $A(s)$ become matrices and Eq.\ \eqref{Gpure} is generalized as
\begin{eqnarray}
\label{analytical:eq19}
G_{ij}(s)&=&\frac{e^{- s}}{2\pi s^2}
\left(\mathsf{L}(s)\cdot \left[\mathsf{I}-
\mathsf{A}(s)\right]^{-1}\right)_{ij},
\end{eqnarray}
where $\mathsf{I}$ is the identity matrix and
\begin{eqnarray}
\label{analytical:eq20}
L_{ij}(s)&=&L_{ij}^\zero+L_{ij}^\one s+L_{ij}^\two s^2,
\end{eqnarray}
\beqa
\label{analytical:eq21}
A_{ij}(s)&=&\rho\frac{x_i}{s^3}\left[(1-e^{-s})L_{ij}(s)-L_{ij}^\zero s\right.\nn
&&\left.+\left(\frac{1}{2}L_{ij}^\zero-L_{ij}^\one\right)s^2\right].
\eeqa
Note that $\lim_{s\to 0} A_{ij}(s)=\text{finite}$, so that $\lim_{s\to 0}s^2 G_{ij}(s)=\text{finite}\neq 0$ by construction. Analogously,
$\lim_{s\rightarrow\infty} e^{ s}G_{ij}(s)=\text{finite}$ also by construction.

The coefficients $L_{ij}^\zero$, $L_{ij}^\one$, and $L_{ij}^\two$ are determined by enforcing the exact conditions \eqref{III.1a}, \eqref{III.1b}, and \eqref{IV.1}. The details of the derivation are presented in Appendix \ref{appB} and here we present the final results. { The} coefficients $L_{ij}^\zero$ and $L_{ij}^\one$ { do not depend upon the first index $i$ and} can be expressed as linear functions of the coefficients {$\{L_{kj}^\two\}$}:
\begin{eqnarray}
\label{analytical:eq30}
L_{ij}^\zero&=&2\pi\frac{1+2\eta}{(1-\eta)^2}-
\frac{12\eta}{1-\eta} \sum_{k} x_k  L_{kj}^\two,
\end{eqnarray}
\begin{eqnarray}
\label{analytical:eq31}
L_{ij}^\one&=&2\pi\frac{1+\eta/2}{(1-\eta)^2}
-\frac{6\eta}{1-\eta} \sum_{k} x_k  L_{kj}^\two,
\end{eqnarray}
{ and} the  coefficients $L_{ij}^\two$ obey the closed set of quadratic equations
\begin{eqnarray}
\label{analytical:eq36}
\frac{L_{ij}^\two}{t_{ij}}&=&
2\pi\frac{1+\eta/2}{(1-\eta)^2}
-\frac{6\eta}{1-\eta} \sum_{k}x_k\left(L_{ik}^\two+L_{kj}^\two\right)\nn
&&
+\frac{6}{\pi}\eta\sum_{k}x_k L_{ik}^\two L_{kj}^\two.
\end{eqnarray}
This closes the problem. Once $L_{ij}^\two$ are known, the contact values are given by
\begin{eqnarray}
\label{analytical:eq34}
\overline{y}_{ij}(1)&=&\frac{L_{ij}^\two}{2\pi t_{ij}}.
\end{eqnarray}

Although here we have taken into account that all the diameters are equal ($\sigma_{ij}=\sigma=1$), the above scheme can be easily generalized to the case of different diameters with the additive rule $\sigma_{ij}=(\sigma_i+\sigma_j)/2$. For symmetric interactions (i.e., $t_{ij}=t_{ji}$) one recovers the PY solution of SHS mixtures for any number of components.\cite{TKR02,Santos98}
It is shown in Appendix \ref{appC} that the pair correlation functions $\overline{g}_{ij}(r)$ derived here are indeed the solution to the PY-like problem posed by Eqs.\ \eqref{analytical:eq16} and \eqref{analytical:eq18}.

\subsection{Case of interest: $t_{11}=t_{22}=t_{21} =0$}
\label{sec3}
In the general scheme described by Eqs.\ \eqref{analytical:eq19}--\eqref{analytical:eq34},  four different
stickiness parameters ($t_{11}$, $t_{12}$, $t_{21}$, and
$t_{22}$) are in principle possible.  With the convention that
in $t_{ij}$ {the particle of species $i$ is always located \emph{below} the
particle of species $j$}, we might consider the simplest possibility of having only one
SHS interaction $t_{12}=t=1/12\tau$ and all other HS interactions
($t_{11}=t_{22}=t_{21} =0$), as illustrated in Fig.\ \ref{fig:fig2}.
This is clearly an intermediate case between a full SHS model ($t_{ij}=t=1/12\tau$)
and a full HS model ($t_{ij}=0$), with a predominance of
repulsive HS interactions with respect to attractive SHS interactions. This is meant to
model the intermediate nature of the original anisotropic Kern--Frenkel potential that
interpolates between a SW and a HS isotropic potentials upon decreasing the coverage, that
is, the fraction of the SW { surface patch with respect to} the full surface of the sphere.

\subsubsection{Structural properties}
If  $t_{11}=t_{22}=t_{21} =0$,  Eq.\ \eqref{analytical:eq36} implies $L_{11}^\two=L_{22}^\two=L_{21}^\two=0$. As a consequence, Eq.\ \eqref{analytical:eq36} for $i=1$ and $j=2$ yields a \emph{linear} equation for $L_{12}^\two$ whose solution is
\beq
L_{12}^\two=2\pi\frac{1+\eta/2}{1-\eta}\frac{t}{1-\eta+6\eta t }.
\label{3.3}
\eeq
 According to Eq.\ \eqref{analytical:eq34},
\beq
\overline{y}_{12}(1)=\frac{1+\eta/2}{(1-\eta)^2}\left(1-\frac{6\eta t}{1-\eta+6\eta t}\right).
\label{3.4}
\eeq
Next, Eqs.\ \eqref{analytical:eq30} and \eqref{analytical:eq31} yield
\beq
\frac{L_{11}^\zero}{2\pi}=\frac{L_{21}^\zero}{2\pi}=\frac{1+2\eta}{(1-\eta)^2},
\label{3.11a}
\eeq
\beq
 \frac{L_{12}^\zero}{2\pi}=\frac{L_{22}^\zero}{2\pi}=\frac{1+2\eta}{(1-\eta)^2}-\frac{12\eta t}{1-\eta}x_1 {\overline{y}_{12}(1)},
\label{3.11b}
\eeq
\beq
\frac{L_{11}^\one}{2\pi}=\frac{L_{21}^\one}{2\pi}=\frac{1+\eta/2}{(1-\eta)^2},
\label{3.13a}
\eeq
\beq
 \frac{L_{12}^\one}{2\pi}=\frac{L_{22}^\one}{2\pi}=\frac{1+\eta/2}{(1-\eta)^2}-\frac{6\eta t}{1-\eta}x_1 {\overline{y}_{12}(1)}.
\label{3.13b}
\eeq

Once the functions $L_{ij}(s)$ are fully determined, Eq.\ \eqref{analytical:eq19} provides the Laplace transforms $G_{ij}(s)$.
{ From Eq.\ \eqref{analytical:eq10}} it follows that $Y_{11}(s)=G_{11}(s)$, $Y_{22}(s)=G_{22}(s)$, $Y_{21}(s)=G_{21}(s)$, and
\beq
Y_{12}(s)=G_{12}(s)-t\overline{y}_{12}(1)e^{- s}.
\label{3.5}
\eeq
{ A numerical inverse Laplace transform} \cite{AW92} allows one to obtain $\overline{g}_{11}(r)$, $\overline{g}_{22}(r)$, $\overline{g}_{21}(r)$, and $\overline{y}_{12}(r)$ for any  packing fraction $\eta$, stickiness parameter $t=1/12\tau$, and mole fraction $x_1$.
In what follows, we will { omit explicit} expressions related to $\overline{g}_{22}(r)$ since they can be readily obtained from $\overline{g}_{11}(r)$ by the exchange $x_1\leftrightarrow x_2$.

The contact values $\overline{g}_{ij}(1^+)=\overline{y}_{ij}(1)$ with  $(i,j)\neq (1,2)$ cannot be obtained from Eq.\ \eqref{analytical:eq34}, unless the associated $t_{ij}$ are first assumed to be nonzero and then the limit $t_{ij}\to 0$ is taken. A more direct method is to realize that, if $t_{ij}=0$,  Eq.\ \eqref{analytical:eq13} gives\beq
\overline{g}_{ij}(1^+)=\lim_{s\to\infty}e^{ s} s G_{ij}(s),\quad (i,j)\neq (1,2).
\label{3.15}
\eeq
The results are
\beq
\overline{g}_{11}(1^+)=\overline{y}_{11}(1)=\frac{1+\eta/2}{(1-\eta)^2}-x_2\frac{6\eta t}{1-\eta}{\overline{y}_{12}(1)},
\label{3.17}
\eeq
\beq
\overline{g}_{21}(1^+)=\overline{y}_{21}(1)=\frac{1+\eta/2}{(1-\eta)^2},
\label{3.17b}
\eeq
\beq
\overline{y}(1)=\frac{1+\eta/2}{(1-\eta)^2}\left(1-x_1x_2\frac{12\eta t}{1-\eta+6\eta t}\right).
\label{global2}
\eeq
It is interesting to note the property $\overline{g}_{11}(1^+)+\overline{g}_{22}(1^+)=\overline{y}_{12}(1)+\overline{g}_{21}(1^+)$.

To obtain the equation of state from the virial route we will need the derivative $\overline{y}_{12}'(1)$. Expanding $e^s G_{12}(s)$ in powers of $s^{-1}$ and using Eq.\ \eqref{analytical:eq13}, one gets
\beqa
\frac{\overline{y}_{12}'(1)}{\overline{y}_{12}(1)}&=&\frac{\eta}{(1-\eta)^2}\left[3t\left(\frac{2-4\eta-7\eta^2}{1+\eta/2}+12 x_1 x_2 \eta\right)\right.\nn
&&\left.-\frac{9}{2}\frac{1-\eta^2}{1+\eta/2}\right].
\label{3.8}
\eeqa

\subsubsection{Thermodynamic properties}

\paragraph{Virial route.}
According to Eq.\ \eqref{analytical:eq6},
\beqa
Z^v&=&1+4\eta\overline{y}(1)-4x_1 x_2\eta t\left[{3\overline{y}_{12}(1)+ \overline{y}_{12}'(1)}\right]\nn
&=&Z_\text{HS}^v-4x_1 x_2 {\eta}t\left[3\frac{1+3\eta}{1-\eta}\overline{y}_{12}(1)+\overline{y}_{12}'(1)\right],\nn
\label{3.27}
\eeqa
where the superscript $v$ denotes the virial route and
\beq
Z_\text{HS}^v=\frac{1+2\eta+3\eta^2}{(1-\eta)^2}
\label{3.26}
\eeq
is the  HS compressibility factor predicted by the virial route in the PY approximation.

\paragraph{Energy route.}
{}From Eq.\ \eqref{analytical:eq7} we have
\beq
\label{3.23}
\frac{u_{\text{ex}}}{\epsilon}=- 12 x_1 x_2 \eta t\overline{y}_{12}(1).
\eeq
The compressibility factor can be obtained from $u_{\text{ex}}$ via the thermodynamic relation \eqref{analytical:eq42}, which in our case reads
\beq
\eta\frac{\partial u_{\text{ex}}/\epsilon}{\partial
\eta}=\frac{1}{\epsilon}\frac{\partial Z}{\partial \beta}=t
\frac{\partial Z}{\partial t}.
\label{3.24}
\eeq
Thus, the compressibility factor derived from the energy route is
\beqa
Z^u&=&Z_\text{HS}^u+\eta\frac{\partial}{\partial\eta}\int_0^t dt'\frac{u_{\text{ex}}(\eta,t')/\epsilon}{t'}\nn
&=&Z_\text{HS}^u-3x_1 x_2 \frac{\eta}{1-\eta}\left[4t\overline{y}_{12}(1)+\frac{\ln\left(1+\frac{6\eta t}{1-\eta}\right)}{1-\eta}\right],\nn
\label{3.25}
\eeqa
where $Z_\text{HS}^u$ plays the role of an integration constant and thus it can be chosen arbitrarily. { It can be shown \cite{S05,S06} that} the energy and the virial routes coincide when the HS system is { the limit of a square-shoulder interaction with} vanishing shoulder width. {}From that point of view one should take $Z_\text{HS}^u=Z_\text{HS}^v$ in Eq.\ \eqref{3.25}. On the other hand, a better description is expected { from} the Carnahan--Starling (CS) equation of state
\beq
Z_\text{HS}^{\text{CS}}=\frac{1+\eta+\eta^2-\eta^3}{(1-\eta)^3}
\label{CS}
\eeq
{ Henceforth} we will take $Z_\text{HS}^u=Z_\text{HS}^{\text{CS}}$.

\paragraph{Compressibility route.}
Expanding $s^2 G_{ij}(s)$ in powers of $s$ it is straightforward to obtain $H_{ij}^\one$ from Eq.\ \eqref{orientational:eq15}. This allows one to use Eqs.\ \eqref{orientational:eq13} and \eqref{orientational:eq17} to get the inverse susceptibility $\chi_T^{-1}$ as
\beq
\chi_T^{-1}=\frac{1+2\eta}{(1-\eta)^4}
\frac{1+2\eta-24x_1x_2 t\eta(1-\eta)\overline{y}_{12}(1)}
{1-x_1x_2\left[\frac{12 t
\eta(1+\eta/2)\overline{y}_{12}(1)}{1+2\eta+36x_1 x_2 t
{\eta^2}\overline{y}_{12}(1)}\right]^2},
\label{3.29}
\eeq
{ that, for an equimolar mixture $(x_1=x_2=\frac{1}{2})$, reduces to}
  \beq
\chi_T^{-1}=\frac{\left[(1-\eta)^2(1+2\eta)+3\eta t\left(2+5\eta-\frac{5}{2}\eta^2\right)\right]^2}{(1-\eta)^5(1-\eta+6\eta t)\left[(1-\eta)^2+3\eta t(4-\eta)\right]}.
\label{3.29b}
\eeq
The associated compressibility factor is then
\beq
Z^c=\frac{1}{\eta}\int_0^\eta d\eta'\, \chi_T^{-1}(\eta').
\label{3.30}
\eeq
The above integral has an analytical solution, but it is too cumbersome to be displayed here.

\subsubsection{Low-density expansion}
\label{subsec_LDE}
In the standard case of SHS mixtures with symmetric coefficients in the potential parameters, the PY closure is known
to reproduce the exact cavity functions to first order in density and thus the  third virial coefficient { (see Appendix \ref{app:appb}). However, this {needs} not be the
case in the RFA description for the present asymmetric case, as further { discussed} below. Note that here, ``exact'' still refers to the simplified problem (orientational average+sticky
limit) of Sections \ref{sec:orientational} and \ref{subsec:sticky}.}

The expansion to first order in density of the Laplace transforms $Y_{ij}(s)$ obtained from Eqs.\ \eqref{analytical:eq10}, \eqref{analytical:eq19}--\eqref{analytical:eq21}, and \eqref{3.3}--\eqref{3.13b} is
\beq
Y_{ij}(s)=e^{-s}\left(s^{-1}+s^{-2}\right)+Y_{ij}^\one(s)\rho+\cdots,
\label{LD1}
\eeq
where the expressions of the first-order coefficients $Y_{ij}^\one(s)$ will be omitted here. Laplace inversion yields
\beq
\overline{y}_{ij}^\one(r)=\left.\overline{y}_{ij}^\one(r)\right|_{\text{exact}}-\Delta \overline{y}_{ij}^\one (r),
\label{LD2}
\eeq
where $\left.y_{ij}^\one(r)\right|_{\text{exact}}$ are the exact first-order functions given by Eqs.\ \eqref{B025}--\eqref{B027} and the deviations $\Delta y_{ij}^\one (r)$ are
\beq
\Delta \overline{y}_{11}^\one (r)=\Theta(2-r)x_2  \frac{2t^2}{r}\cos^{-1}\frac{r}{2},
\label{LD3}
\eeq
\beq
\Delta \overline{y}_{12}^\one (r)=\Theta(2-r)  t\left(2\sqrt{1-r^2/4}-r \cos^{-1}\frac{r}{2}\right),
\label{LD4}
\eeq
\beq
\Delta \overline{y}_{21}^\one (r)=-\Delta \overline{y}_{12}^\one (r).
\label{LD5}
\eeq
In the case of the global quantity $\overline{y}^\one(r)$ the result is
\beq
\overline{y}^\one(r)=\left.\overline{y}^\one(r)\right|_{\text{exact}}-\Delta \overline{y}^\one (r),
\label{LD6}
\eeq
where $\left.\overline{y}^\one(r)\right|_{\text{exact}}$ is given by Eq.\ \eqref{B025b} and
\beq
\Delta \overline{y}^\one (r)=\Theta(2-r)x_1 x_2  \frac{2t^2}{r}\cos^{-1}\frac{r}{2}.
\label{LD7}
\eeq
While the main qualitative features { of the exact cavity function} are preserved, there exist quantitative differences. The first-order functions $\overline{y}_{11}^\one(r)$, $\overline{y}_{22}^\one(r)$, and $\overline{y}^\one(r)$ predicted by the RFA account for the exact coefficient of $t$ but do not include the exact term of order $t^2$ proportional to $r^{-1}\cos^{-1}(r/2)$. In the case of $\overline{y}_{12}^\one(r)$ and $\overline{y}_{21}^\one(r)$ the exact term of order $t$ proportional to $2\sqrt{1-r^2/4}-r\cos^{-1}(r/2)$ is lacking.
Also, while the combination $\overline{y}_{11}^\one(r)+\overline{y}_{22}^\one(r)-\overline{y}_{12}^\one(r)-\overline{y}_{21}^\one(r)$ vanishes in the RFA,  the exact result is proportional to $t^2 r^{-1}\cos^{-1}(r/2)$.
In  short, {the RFA} correctly accounts for the polynomial terms in $\left.y_{ij}^\one(r)\right|_{\text{exact}}$ but { misses} the non-polynomial terms.

As for the thermodynamic quantities, expansion of Eqs.\ \eqref{3.27}, \eqref{3.25}, and \eqref{3.30} gives
\beqa
Z^v&=&1+4\left(1-3{x_1x_2} t\right)\eta+10\left[1-6{x_1x_2}t\left(1-\frac{4}{5} t\right)\right]\eta^2\nn
&&+\mathcal{O}(\eta^3),
\label{3.33}
\eeqa
\beqa
Z^u&=&1+4\left(1-3{x_1x_2} t\right)\eta+10\left[1-6{x_1x_2}t\left(1-\frac{6}{5} t\right)\right]\eta^2
\nn
&&+\mathcal{O}(\eta^3),
\label{3.32}
\eeqa
\beqa
Z^c&=&1+4\left(1-3{x_1x_2} t\right)\eta+10\left[1-6{x_1x_2}t\left(1-\frac{8}{5} t\right)\right]\eta^2
\nn
&&+\mathcal{O}(\eta^3).
\label{3.34}
\eeqa
Comparison with the exact third virial coefficient, Eq.\ \eqref{B031}, shows that the coefficient of $t^2$ is not { correct, with the exact factor $4-3\sqrt{3}/\pi\simeq 2.35$ replaced by} $2$, $3$, and $4$ in Eqs.\ \eqref{3.33}--\eqref{3.34}, respectively. { One consequence is that} the virial and energy routes predict the third virial coefficient much better than the compressibility route.
{ A possible improvement is through the} interpolation formula
\beq
Z^{v,u}=\alpha\left(Z^v+Z_\hs^{\text{CS}}-Z_\hs^v\right)+(1-\alpha)Z^u,
\label{Zvu}
\eeq
where $\alpha=3\sqrt{3}/\pi-1\simeq 0.65$ { with the proviso that} $Z_\hs^u=Z_\hs^{\text{CS}}$ in Eq.\ \eqref{3.25}.
Equation \eqref{Zvu} { then} reduces to the CS equation of state if $t=0$ and { reproduces the exact third virial coefficient when}
$t\neq 0$.
\subsubsection{Phase transition and critical point}
\label{subsec:phase}
In the limit of isotropic interaction ($t_{ij}=t$), our model reduces to the usual SHS Baxter adhesive
one-component model. In spite of the fact that the model is, strictly speaking, known to be pathological,\cite{Stell91} it displays a critical behavior that was numerically studied in some details by MC techniques.\cite{Miller03,Miller04}
The corresponding binary mixture also displays well defined critical properties that, interestingly, are even {free from} any pathological behavior.
\cite{Zaccarelli00}
{Moreover, the mechanism behind the pathology of the isotropic Baxter model hinges crucially on the geometry of certain close-packed clusters involving 12 or more equal-sized spheres.\cite{Stell91} On the other hand, our Janus model, having frozen orientations, cannot sustain those pathological configurations. }

Within the PY approximation, the critical behavior of the original one-component Baxter SHS  model was studied using the
compressibility and virial routes,\cite{Baxter68} as well as the energy route,\cite{Watts71} in the latter case with the { implicit assumption} $Z_\hs^u=Z_\hs^\cs$.
Numerical simulations indicate that the critical point found through the energy route { is the closest to numerical simulation results}.\cite{Miller03,Miller04}

As the present specific model ({ with}, $t_{ij}=t\delta_{i1}\delta_{j2}$) is, in some sense, intermediate between the fully isotropic Baxter SHS one-component model (that has a full, albeit peculiar,
gas-liquid transition) and the equally isotropic HS model (that, lacking any attractive part in the potential, cannot have any gas-liquid
transition), it is { then interesting to ask whether in the equimolar case {($x_1=x_2=\frac{1}{2}$)} {it} still presents a critical gas-liquid transition}.

\begin{table}
\caption{Location of the critical point in the RFA, according to different routes.}
\label{tab1}
\begin{ruledtabular}
\begin{tabular} {cccc}
Route&$\tau_c$&$\eta_c$&$Z_c$\\
\hline
virial, Eq.\ \protect\eqref{3.27}&$0.02050$&$0.1941$&$0.3685$\\
energy, Eq.\ \protect\eqref{3.25}&$0.0008606$&$0.2779$&$0.2906$\\
hybrid virial-energy, Eq.\ \protect\eqref{Zvu}&$0.01504$&$0.1878$&$0.3441$\\
\end{tabular}
\end{ruledtabular}
\end{table}

The answer depends on the route followed to obtain the pressure. As { seen} from Eq.\ \eqref{3.29b}, the compressibility route yields a positive definite $\chi_T^{-1}$, so that no critical point is predicted by this route. On the other hand, an analysis of the virial [Eq.\ \eqref{3.27}], energy [Eq.\ \eqref{3.25} with $Z_\hs^u=Z_\hs^\cs$], and hybrid virial-energy [Eq.\ \eqref{Zvu}] equations of state reveals the existence of van der Waals loops with the respective critical points shown in Table \ref{tab1}. The energy route predicts a critical value $\tau_c$ about {twenty times smaller than} the values predicted by the other two routes.

\begin{figure}[htbp]
\centering
\includegraphics[width=8.5cm]{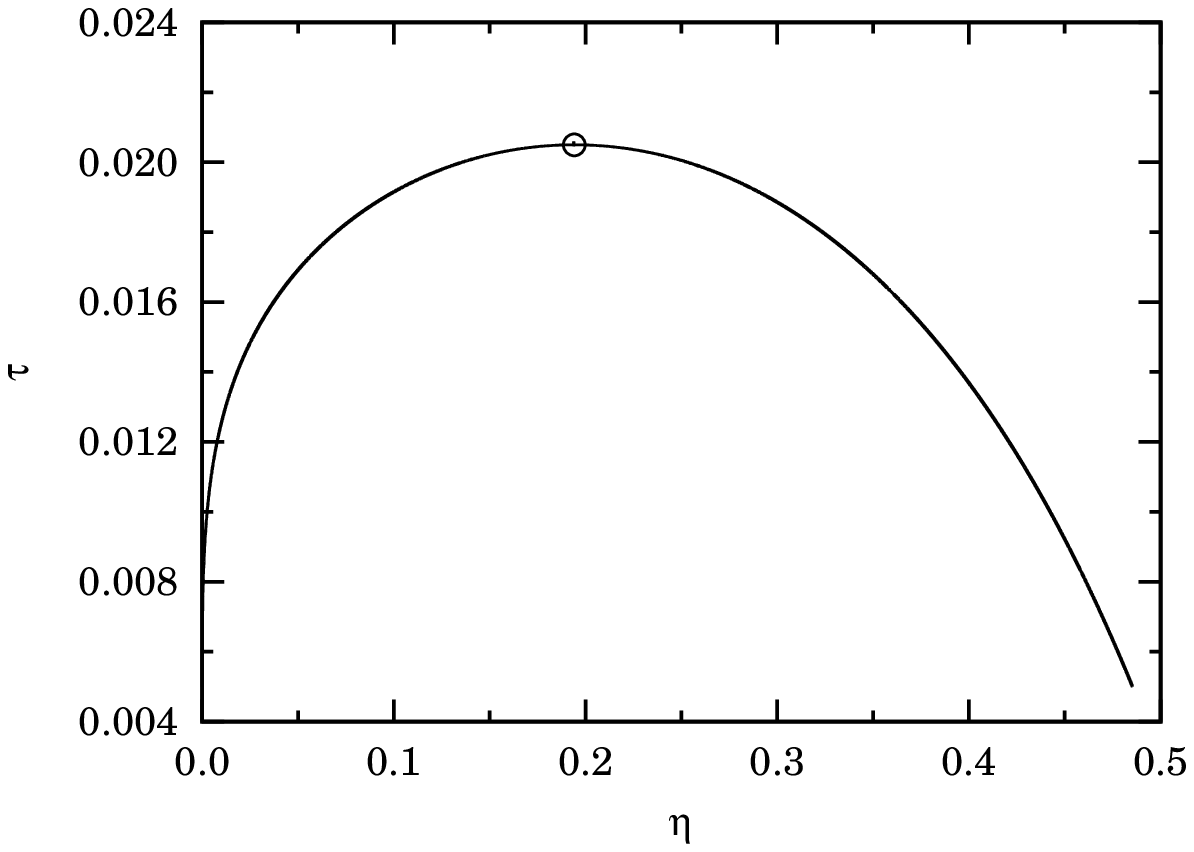}\\
\includegraphics[width=8.5cm]{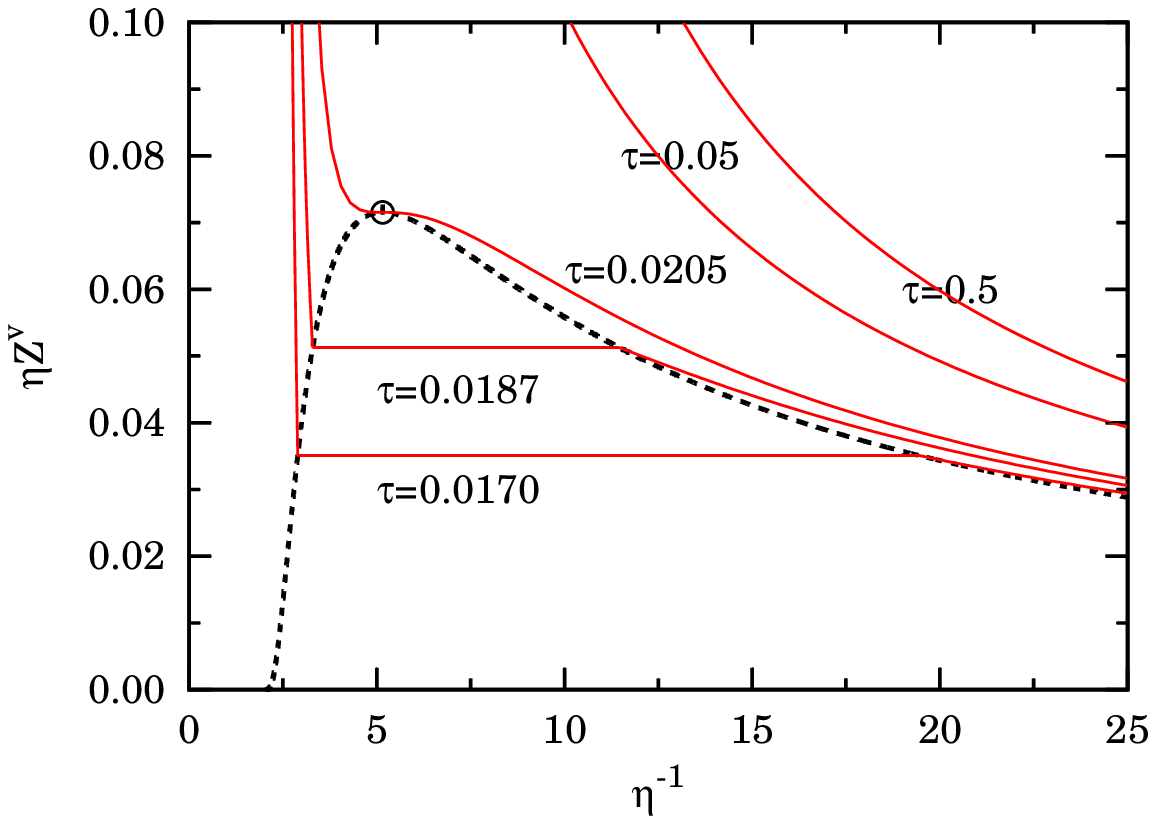}
\caption{Binodals from the RFA virial route in the equimolar
  $x_1=\frac{1}{2}$ case. The phase diagram is depicted in the $(\eta,\tau)$ plane ({solid line,} top panel) and
  in the $(\eta^{-1},\eta Z^v)$ plane ({dashed line, }bottom panel). {A few characteristic isotherms are plotted in the bottom panel}. The critical point is found at
  $\eta_c \simeq 0.1941$, $\tau_c \simeq 0.02050$, {and $\eta_c Z_c\simeq 0.07153$} ({indicated by a circle in both panels)}.}
\label{fig:fig8}
\end{figure}

As an illustration, Fig.\ \ref{fig:fig8} shows the binodal and a few isotherms, as obtained from the virial route.

\subsubsection{A modified approximation}
The failure of the RFA  to reproduce the exact cavity functions to first order in density (and hence the third virial coefficient) for asymmetric interactions ($t_{ij}\neq t_{ji}$) reveals the { price paid for using the orientationally averaged quantities} $\overline{g}_{ij}(r)$ instead of the true pair correlation functions ${g}_{ij}(\rr)$.

A simple way { of getting around this drawback} for sufficiently low values of both $\eta$ and $t$ consists of modifying the RFA as follows:
\beq
\overline{y}_{ij}(r)\to \overline{y}_{ij}(r)+\Delta \overline{y}_{ij}^\one(r)\rho,
\label{M1}
\eeq
where the functions $\Delta \overline{y}_{ij}^\one(r)$ are given by Eqs.\ \eqref{LD3}--\eqref{LD5}.
We will refer to this as the \emph{modified} { rational-function} approximation (mRFA).
Note that Eq.\ \eqref{M1} implies that $\overline{g}_{ij}(r)\to \overline{g}_{ij}(r)+\Delta \overline{y}_{ij}^\one(r)\rho$, except if $(i,j)=(1,2)$, in which case
$
\overline{g}_{12}(r)\to \overline{g}_{12}(r)+\Delta \overline{y}_{12}^\one(r)\rho+\Delta \overline{y}_{12}^\one(1)\delta_+(r-1)\rho t
$.

Since the extra terms in Eq.\ \eqref{M1} are proportional to $t$ or $t^2$, this modification can produce poor results { for sufficiently large stickiness (say, $t\gtrsim 1$) as, for instance, near the critical point.}

\section{Numerical calculations}
\label{sec:numerical}
\subsection{Details of the simulations}
\label{subsec:details}
In order to check the theoretical predictions previously reported, we have
performed NVT (isochoric-isothermal)  MC simulations using the Kern--Frenkel potential
defined in Eqs.\ \eqref{mapping:eq1}--\eqref{mapping:eq4} with a single attractive SW patch
(green in the color code of Fig.\ \ref{fig:fig1}) covering one of the two hemispheres, { and with up-down symmetry as depicted
in Fig.\ \ref{fig:fig2}.
Particles are then not allowed to rotate around but only to translate rigidly.}

The model is completely defined by specifying the relative width $\lambda-1$, the
concentration of one species (mole fraction) $x_1=1-x_2$,  the reduced density $\rho^*=\rho\sigma^3$,  and the reduced temperature $T^*=k_BT/\epsilon$.

 In order to make sensible comparison with the RFA theoretical predictions,
we have selected the value $\lambda-1=0.05$ as a well width, which is known to be well represented by the SHS limit, \cite{Malijevsky06}
and use Baxter's temperature parameter $\tau=\left[4(\lambda^3-1)\left(e^{1/T^*}-1\right)\right]^{-1}$ [see Eq.\ \eqref{analytical:eq2}] instead of
$T^{*}$. {It is interesting to note that, while the unconventional phase diagram found in the simulations of Ref.\ \onlinecite{Sciortino10} corresponded to a larger well width ($\lambda=1.5$),
the value $\lambda=1.05$ is in fact closer to the experimental conditions of Ref.\ \onlinecite{Hong08}.}

During the
simulations we have computed the orientational averaged pair { correlation functions defined by Eqs.\ \eqref{orientational:eq3a} and \eqref{orientational:eq3b}, { accumulating} separate histograms when $z_2-z_1>0$ or   $z_1-z_2>0$ in order to distinguish between functions $\overline{g}_{12}(r)=g_{12}^+(r)$
and $\overline{g}_{21}(r)=g_{12}^-(r)$.}

The compressibility
factor $Z=\beta P/\rho$ has been evaluated from the values of $\overline{y}_{ij}(r)$ at $r=\sigma$ and $r=\lambda\sigma$ by following Eq.\ \eqref{analytical:eq1} with $t_{ij}=(12\tau)^{-1}\delta_{i1}\delta_{j2}${, and  the reduced excess internal energy
per particle $u^*_\ex=u_\ex/\epsilon$ has been evaluated directly from simulations.}

In all our simulations, we used $N = 500$ particles, periodic boundary conditions, an equilibration time of around $10^5$ MC steps (where a MC step corresponds to a single particle displacement), and a production time of about
$10^8$ MC steps for the structure calculations and up to $5\times 10^8$ MC steps for the thermophysical calculations. The maximum particle displacement was determined during the first stage of the equilibration run in such a way as to ensure an average acceptance ratio of 50\% at production time.

\subsection{Results for non-equimolar binary mixtures}
\label{subsec:results_ne}
As a preliminary attempt, we consider a binary mixture under non-equimolar conditions, { to}
avoid possible pathologies arising from the symmetry of the two components akin to
those occurring in ionic systems. As we shall see below, no such pathologies are found.

In the present case, we consider a system with $x_1=1/5$ and $x_2=1-x_1=4/5$, so that
the majority of the spheres have (green) attractive patches pointing {in the direction of} $-\widehat{\mathbf{z}}$.

\begin{figure}[htbp]
\centering
\includegraphics[width=8.5cm]{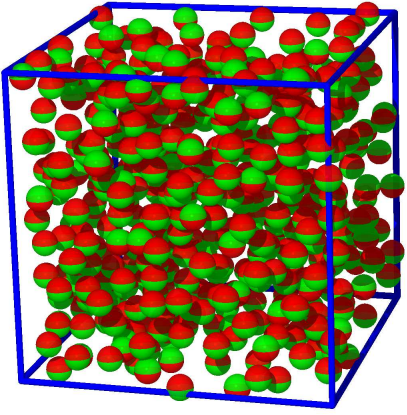}
\caption{Snapshot of an equilibrated MC simulation under non-equimolar conditions
  ($x_1=1/5$)  with Baxter temperature $\tau=0.1$
  and density $\rho^*=0.3$. In the simulations
  we used a total number of particles $N=500$.}
\label{fig:fig3}
\end{figure}

A snapshot of an equilibrated configuration is shown in Fig.\ \ref{fig:fig3}. This configuration was obtained
using $N=500$ particles at $\rho^*=0.3$ and Baxter temperature $\tau=0.1$ (corresponding to $T^*\simeq 0.354$).
Note that the above chosen state point  ($\rho^*=0.3$ and $\tau=0.1$) lies well inside the critical region of the full
Baxter SHS adhesive model as obtained from direct MC simulations,\cite{Miller03, Miller04} although of course the present case
is expected to display a different behavior { as} only a fraction of about $x_1 x_2=4/25$ of the pair contacts are attractive.

\begin{figure}
  \includegraphics[width=8.5cm]{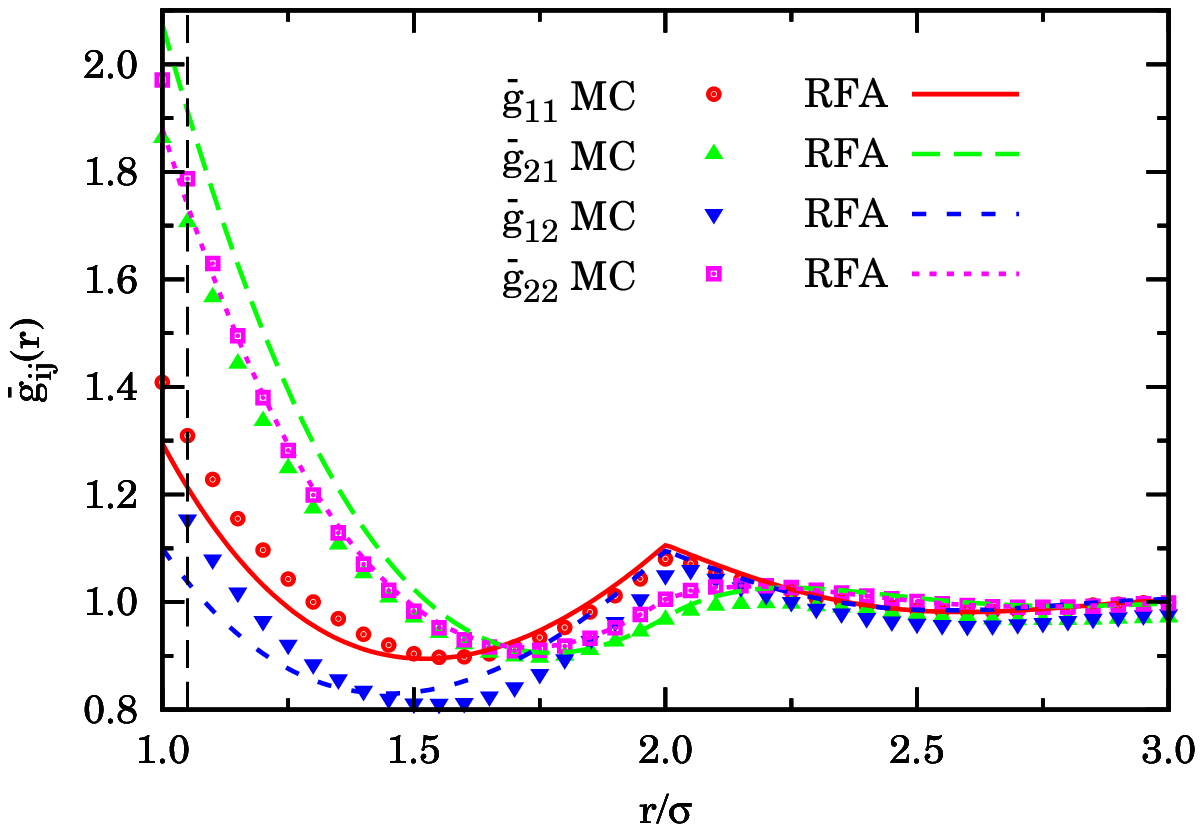}\\
  \includegraphics[width=8.5cm]{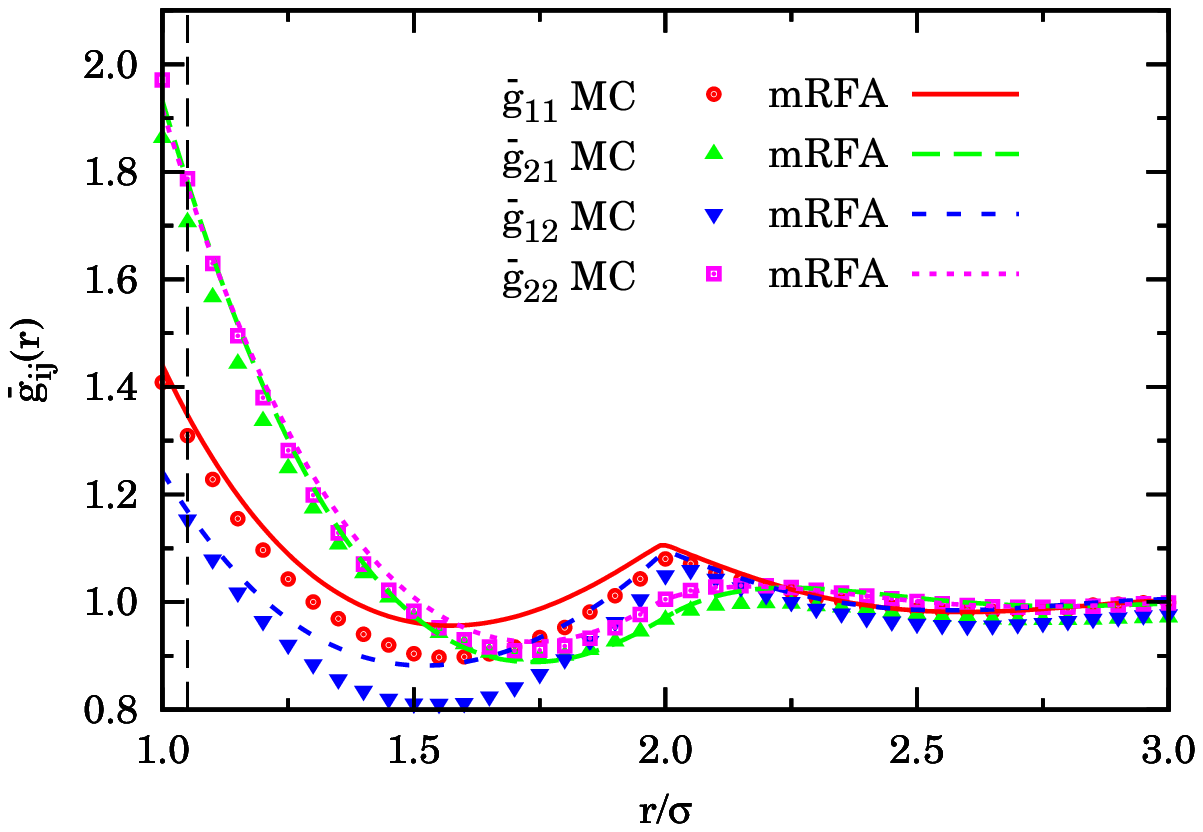}
\caption{Comparison between MC simulations and the theoretical predictions from RFA (top) and mRFA (bottom)  for the orientational averaged distribution functions $\overline{g}_{11}(r)$, $\overline{y}_{12}(r)$, $\overline{g}_{21}(r)$,
and $\overline{g}_{22}(r)$ under non-equimolar conditions ($x_1=1/5$) at density $\rho^*=0.5$ and Baxter temperature $\tau=0.2$. The dashed vertical line indicates the range $r=\lambda=1.05$ of the $(1,2)$ SW interaction used in the simulations.
Note that  the radial distribution function $\overline{g}_{12}(r)$ is obtained in the MC case by multiplying $\overline{y}_{12}(r)$ in the region $1\leq r\leq \lambda=1.05$ by the factor $e^{1/T^*}\simeq 8.93$; in the theoretical cases (SHS limit)  $\overline{g}_{12}(r)$ is obtained by adding the singular term $(12\tau)^{-1}\overline{y}_{12}(1)\delta_+(r-1)$ to $\overline{y}_{12}(r)$. The error bars on the MC data are within the size of the symbols used.
} \label{fig:fig4}
\end{figure}

A good insight on the structural properties of the system can be obtained from the computation of the radial distribution functions
$\overline{g}_{11}(r)$, $g_{12}^{+}(r)=\overline{g}_{12}(r)$, $g_{12}^{-}(r)=\overline{g}_{21}(r)$, and $\overline{g}_{22}(r)$. This is reported  in Fig.\ \ref{fig:fig4} for
a state point at density $\rho^*=0.5$ and Baxter temperature $\tau=0.2$ (corresponding to $T^*\simeq 0.457$). Note that in the case of the pair $(1,2)$ what is actually plotted is the cavity function $\overline{y}_{12}(r)$ rather than $\overline{g}_{12}(r)$, as explained in the caption of Fig.\ \ref{fig:fig4}.

The relatively low value $\tau=0.2$ gives rise to clearly distinct features of the four MC functions $\overline{g}_{ij}(r)$ (which would collapse to a common HS distribution function in the high-temperature limit $\tau\to\infty$). We observe that $\overline{g}_{22}(r)\simeq \overline{g}_{21}(r)>\overline{g}_{11}(r)>\overline{y}_{12}(r)$ in the region $1\leq r\lesssim 1.5$. Moreover, $\overline{g}_{11}(r)$ and  $\overline{g}_{12}(r)$ exhibit a rapid change around $r=2$. This is because when a pair $(1,1)$ is separated a distance $r\approx 2$ there is enough room to fit a particle of species 2 in between and that particle will interact attractively with the particle of the pair $(1,1)$ below it. In the case of the pair $(1,2)$  separated a distance $r\approx 2$, the intermediate particle can be either of species 1 (interacting attractively with the particle of species 2 above it) or of species 2 (interacting attractively with the particle of species 1 below it). The same argument applies to a pair $(2,2)$ separated a distance $r\approx 2$, but in that case the intermediate particle must be of species 1 to produce an attractive interaction; since the concentration of species 1 is four times smaller than that of species 2, the rapid change of $\overline{g}_{22}(r)$ around $r=2$ is much less apparent than that of $\overline{g}_{11}(r)$ and  $\overline{g}_{12}(r)$ in Fig.\ \ref{fig:fig4}. On the other hand, in a pair $(2,1)$ separated a distance $r\approx 2$ an intermediate particle of either  species 1 or of species 2 does not create any attraction and thus $\overline{g}_{21}(r)$ is rather smooth at $r=2$.
{ In short,} the pair correlation function $\overline{g}_{21}(r)$ exhibits HS-like features, $\overline{g}_{12}(r)$ exhibits SW-like features (very high values in the region $1\leq r\leq \lambda$ and discontinuity at $r=\lambda$ due to the direct SW interaction; rapid change around $r=2$ due to indirect SW interaction), while $\overline{g}_{11}(r)$ and $\overline{g}_{22}(r)$ exhibit intermediate features (rapid change around $r=2$ due to indirect SW interaction).

It is { rewarding} to notice how well the MC results are reproduced at a semi-quantitative level by the RFA theory (top panel of Fig.\ \ref{fig:fig4}), in spite of the various approximations involved. In this respect, it is worth
recalling that while MC simulations deal with the real Kern--Frenkel potential, albeit with constrained angular orientations, the RFA theory
deals with the asymmetric binary mixture resulting from the mapping described in Section \ref{sec:mapping}, and this represents an indirect test
of the correctness of the procedure. In addition, the RFA does not attempt to describe the true SW interaction (i.e., finite $\lambda-1$ and $T^*$) but the SHS limit ($\lambda-1\to 0$ and $T^*\to 0$ with finite $\tau$). This limit replaces the high jump of $\overline{g}_{12}(r)$ in the region $1\leq r\leq \lambda$ by a Dirac's delta at $r=1^+$ and the rapid change of $\overline{g}_{12}(r)$, $\overline{g}_{11}(r)$, and $\overline{g}_{22}(r)$ around $r=2$ by a kink. { Finally,} the RFA worked out in Sec.\ \ref{subsec:rfa} results from a heuristic generalization to asymmetric mixtures ($\tau_{ij}\neq \tau_{ji}$) of the PY exact solution for SHS symmetric mixtures ($\tau_{ij}= \tau_{ji}$),\cite{PS75,Barboy75,Barboy79,Zaccarelli00,TKR02,Santos98} but it is not the solution of the PY theory for the asymmetric problem, as discussed in Sec.\ \ref{subsec:simple}. As a matter of fact, the top panel of Fig.\ \ref{fig:fig4} shows that some of the drawbacks of the RFA observed to first order in density in Sec.\ \ref{subsec_LDE} [see Eqs.\ \eqref{LD2}--\eqref{LD5}] remain at finite density: in the region $1\leq r\lesssim 1.5$ the RFA underestimates $\overline{y}_{12}(r)$, $\overline{g}_{11}(r)$, and $\overline{g}_{22}(r)$, while it overestimates $\overline{g}_{21}(r)$.
These discrepancies are widely remedied, at least in the region $1\leq r\lesssim 1.25$, by the mRFA approach [see Eq.\ \eqref{M1}], as shown in the { bottom} panel of Fig.\ \ref{fig:fig4}. In particular, the contact values are well accounted for by the mRFA, as well as the property $\overline{g}_{22}(r)\simeq \overline{g}_{21}(r)$. We have observed that the limitations of the correlation functions $\overline{g}_{ij}(r)$ predicted by the RFA become more important as the density and, especially, the stickiness increase and in those cases the mRFA version does not help much since the correction terms $\Delta\overline{y}_{ij}^\one(r)\rho$, being proportional to $\rho$ and to $t$ or $t^2$, become too large.

\begin{figure}[htbp]
   \centering
   \includegraphics[width=8.5cm]{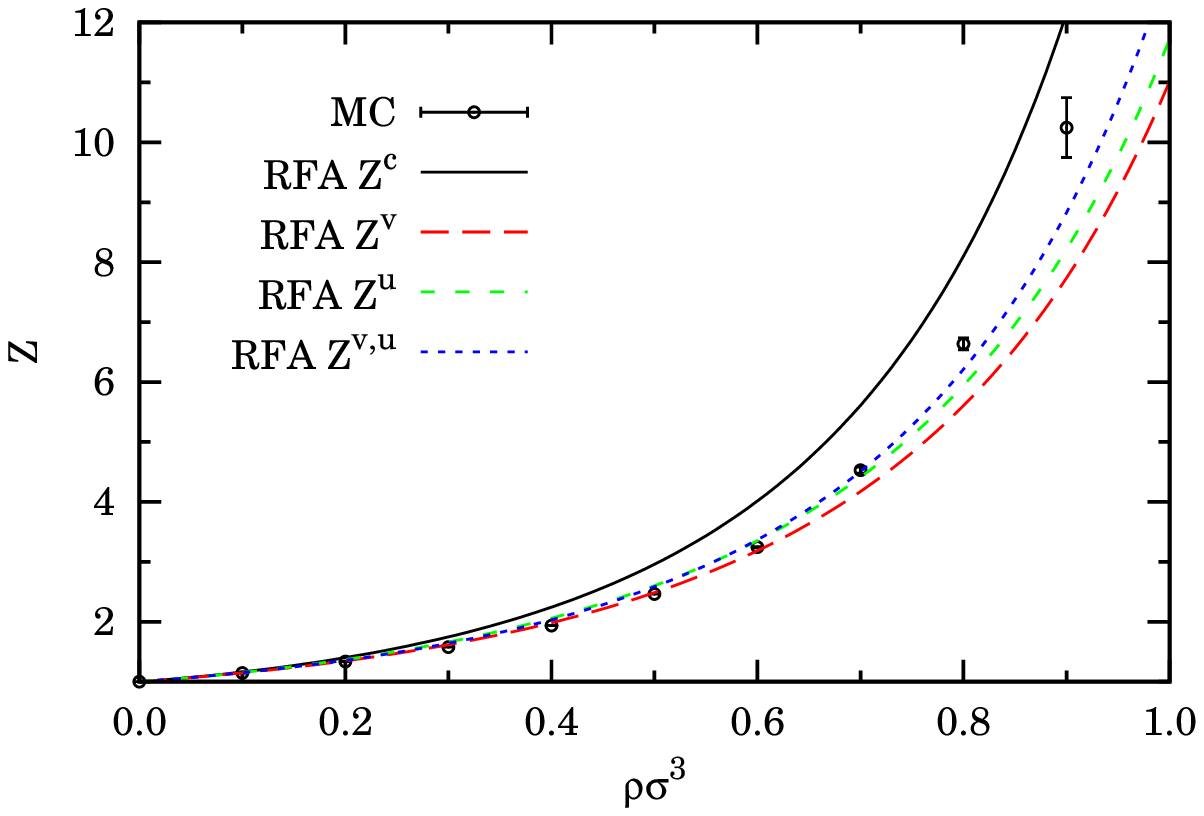} \\
    \includegraphics[width=8.5cm]{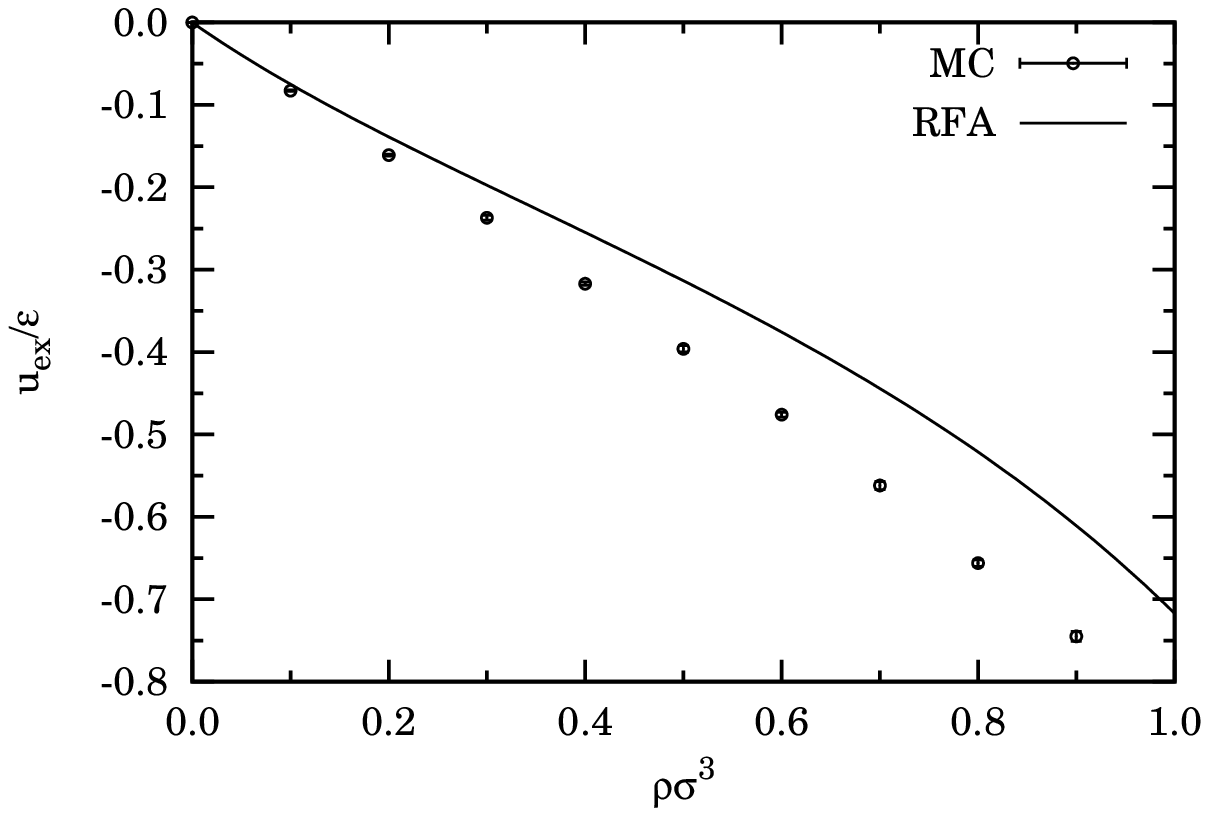}
   \caption{Comparison of MC simulations and RFA theory for the thermodynamics. Both the compressibility factor $Z=\beta P/\rho$ (top)
and the excess internal energy per particle $u_{\ex}/\epsilon$ (bottom) are displayed as  functions of density
for the non-equimolar case $x_1=1/5$ and for Baxter temperature $\tau=0.1$. In the case of the compressibility factor (top),
results for all four routes (compressibility, virial, energy, and hybrid virial-energy) are reported.} \label{fig:fig5}
\end{figure}
Next we consider thermodynamic quantities, as represented by the compressibility factor $Z=\beta P/\rho$ and the
excess internal energy per particle $u_{\ex}/\epsilon$, both directly accessible from NVT numerical MC simulations.
These quantities are depicted in Fig.\ \ref{fig:fig5} as functions of the reduced density $\rho^*$ and for a Baxter temperature $\tau=0.1$.
In both cases, the results for the RFA theory are also included. In the case of the compressibility factor, all four routes are displayed: compressibility [Eqs.\ \eqref{3.4}, \eqref{3.29}, and \eqref{3.30}], virial [Eqs.\ \eqref{3.4}, \eqref{3.8}, and \eqref{3.27}],  energy [Eq.\ \eqref{3.4} and \eqref{3.25} with  $Z_\text{HS}^u=Z_\text{HS}^{\text{CS}}$], and hybrid virial-energy [Eq.\ \eqref{Zvu}]. In the case of $u_\ex/\epsilon$, only the genuine energy route, Eq.\ \eqref{3.23}, is considered. { Note that all RFA} thermodynamic quantities, including Eq.\ \eqref{3.30}, have explicit analytical expressions.

The top panel of Fig.\ \ref{fig:fig5} shows that up to $\rho^*\approx 0.7$ the MC data for the compressibility factor are well predicted by the theoretical $Z^v$ and, especially, $Z^u$  and $Z^{v,u}$. Beyond that point, the  numerical results are bracketed by the compressibility route, that overestimates the pressure, and the hybrid virial-energy route, that on the contrary underestimates it.
It is interesting to note that, while $Z^v<Z^{v,u}<Z^u$ to second order in density [cf.\ Eqs.\ \eqref{3.33}, \eqref{3.32}, and \eqref{Zvu}], the difference $Z^v-Z^v_\hs$ grows with density more rapidly than the difference $Z^u-Z^u_\hs$ and so both quantities cross at a certain density ($\rho^*\simeq 0.567$ if $x_1=1/5$ and $\tau=0.1$). Therefore, even though $Z^v<Z^u$ (because $Z_\hs^v<Z_\hs^\cs$), $Z^{v,u}$ is no longer bracketed by $Z^v$ and $Z^u$ beyond that density ($\rho^*\simeq 0.567$ in the case of Fig.\ \ref{fig:fig5}).
On balance, the  virial-energy  route appears to be the most effective one in reproducing the numerical simulations results of the pressure at $x_1=1/5$ and $\tau=0.1$.

As for the internal energy, the bottom panel of Fig.\ \ref{fig:fig5} shows that the RFA underestimates its magnitude as a direct consequence of the {underestimation} of the contact value $\overline{y}_{12}(1)$ [see Eq.\ \eqref{3.23}]. Although not shown in Fig.\ \ref{fig:fig5}, we have checked that the internal energy per particle obtained from the virial equation of state \eqref{3.27} via the thermodynamic relation \eqref{3.24} exhibits a better agreement with the simulation data than the direct energy route.

\subsection{Results for equimolar binary mixtures}
\label{subsec:results}

Having rationalized the non-equimolar case, the equimolar ($x_1=x_2=1/2$) case can now be safely tackled. The equimolarity condition makes the system {be} more akin to the original Janus model (see Fig.\ \ref{fig:fig1}) since both spin orientations are equally represented.

\begin{figure}[htbp]
\centering
\includegraphics[width=8.5cm]{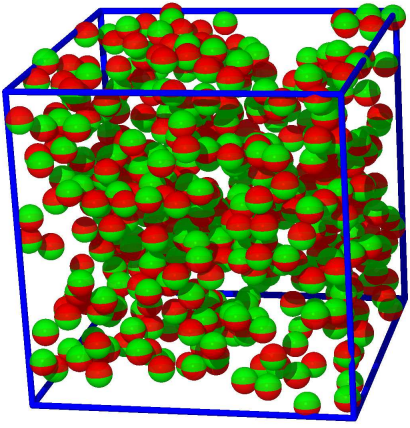}
\caption{Same as in Fig.\ \protect\ref{fig:fig3}, but for an equimolar mixture ($x_1=x_2=1/2$).}
\label{fig:fig6}
\end{figure}

\begin{figure}
  \includegraphics[width=8.5cm]{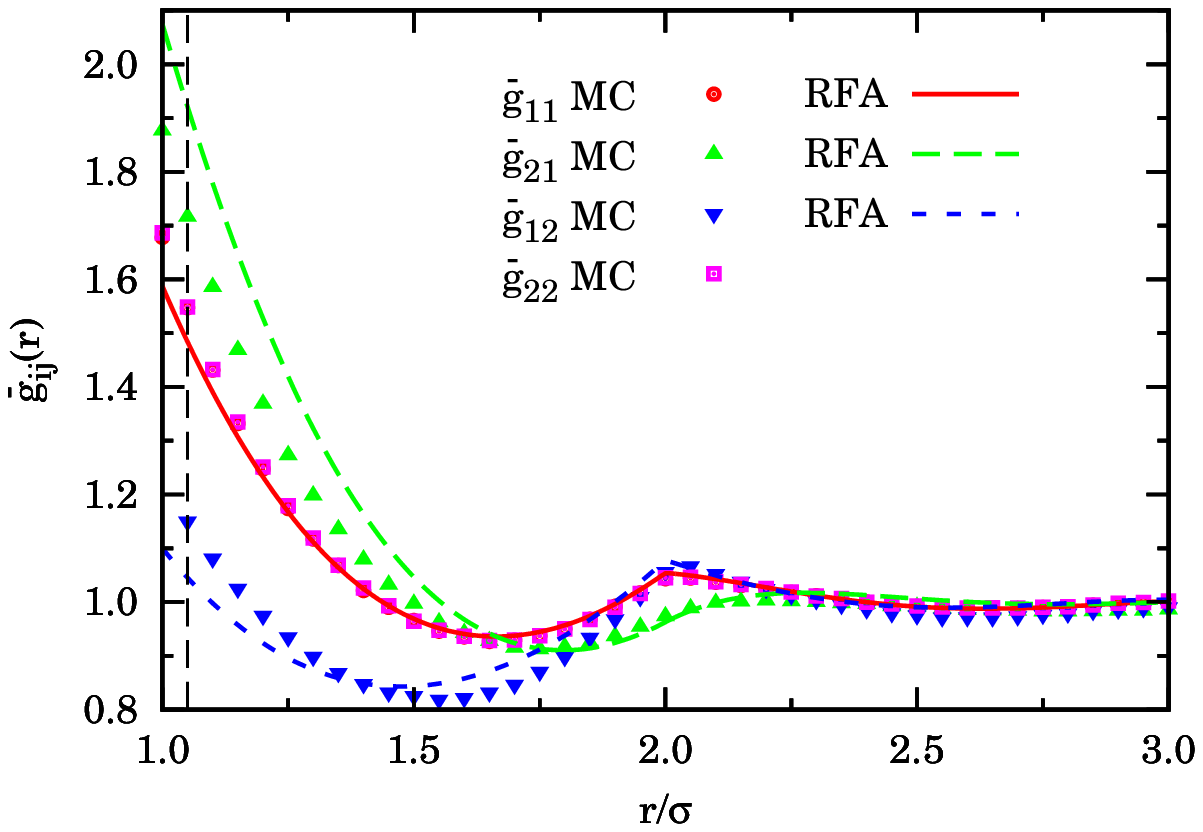}\\
  \includegraphics[width=8.5cm]{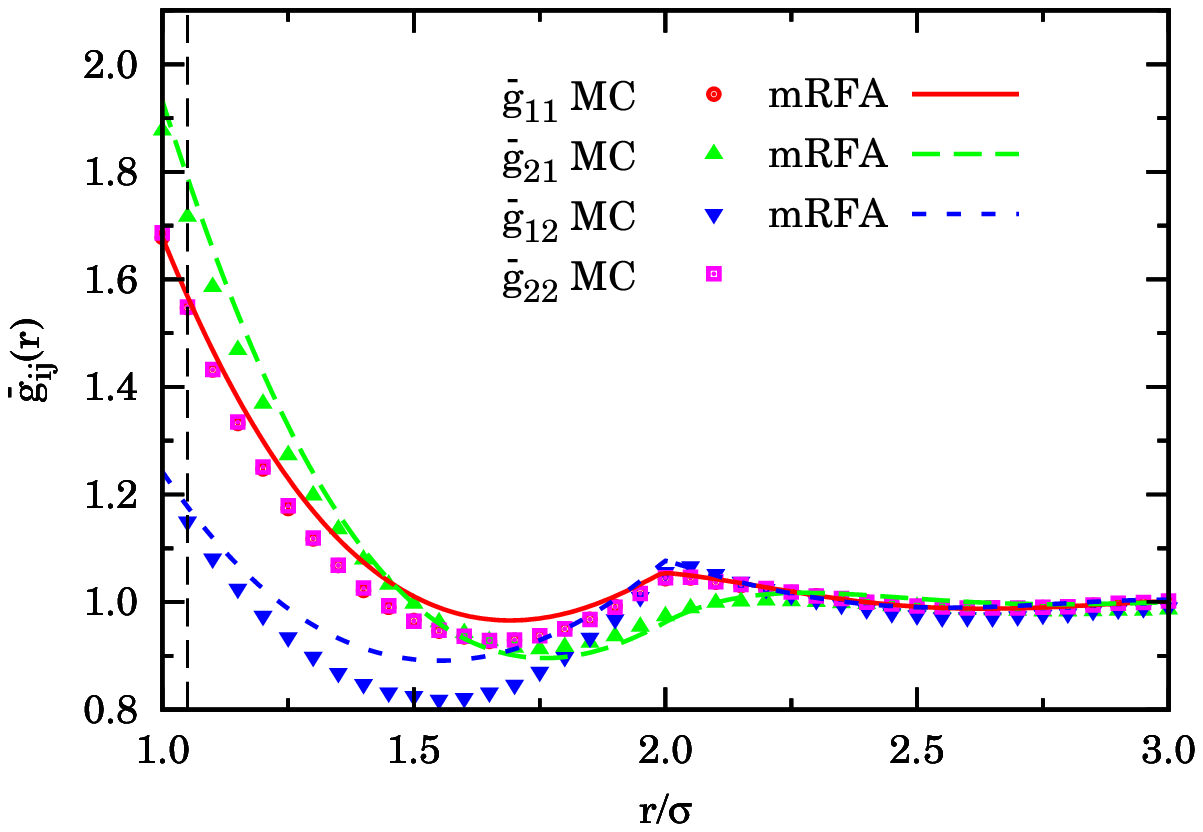}
\caption{Same as in Fig.\ \protect\ref{fig:fig4}, but for an equimolar mixture ($x_1=x_2=1/2$).
} \label{fig:fig7}
\end{figure}

We start with the snapshot  of an equilibrated configuration at density $\rho^*=0.3$ and Baxter temperature
$\tau=0.1$, { that {are} the same values used in the non-equimolar case}. {}From Fig.\ \ref{fig:fig6} it can be visually inspected that, in contrast to the non-equimolar case of Fig.\ \ref{fig:fig3}, the number of particles with spin up { matches that} with spin down.
This equimolar condition then facilitates the interpretation of the corresponding structural properties, as
illustrated by the radial distribution function $\overline{g}_{ij}(r)$ given in Fig.\ \ref{fig:fig7}.

This was obtained at a Baxter temperature $\tau=0.2$ and a density $\rho^*=0.5$, a state point that is expected to be
outside the coexistence curve (see below), but inside the liquid region. { Again,} this is the same state point
as the non-equimolar case previously discussed. {
Now $\overline{g}_{11}(r)=\overline{g}_{22}(r)$ (independently computed) as it should. Notice that} the main features commented before in connection with Fig.\ \ref{fig:fig4} persist. In particular, $\overline{g}_{21}(r)>\overline{g}_{11}(r)=\overline{g}_{22}(r)>\overline{y}_{12}(r)$ in the region $1\leq r\lesssim 1.5$,
$\overline{g}_{11}(r)=\overline{g}_{22}(r)$ and $\overline{g}_{12}(r)$ present  rapid changes around $r=2$, and $\overline{g}_{21}(r)$ exhibits a HS-like shape. Also, as before, the RFA captures quite well the behaviors of the correlation functions (especially noteworthy in the case of $\overline{g}_{11}=\overline{g}_{22}$).
{On} the other hand, the RFA tends to underestimate $\overline{y}_{12}(r)$ and $\overline{g}_{11}(r)=\overline{g}_{22}(r)$ and to overestimate $\overline{g}_{21}(r)$ in the region $1\leq r\lesssim 1.5$. The use of the modified version (mRFA)  partially corrects those discrepancies near contact, although the general behavior only improves in the case of $\overline{g}_{21}(r)$.

Comparison between Figs.\ \ref{fig:fig4} and \ref{fig:fig7} shows that $\overline{y}_{12}(r)$ and $\overline{g}_{21}(r)$ are very weakly affected by the change in composition. In fact, the spatial correlations between particles of species 1 and 2
mediated by a third particle (i.e., to first order in density) {depend} strongly on which particle (1 or 2) is above o
below the other one but not on the nature of the third intermediate particle, as made explicit by Eqs.\  \eqref{B026} and \eqref{B027}. Of course, higher-order terms (i.e., two or more intermediate particles) create a composition-dependence on $\overline{y}_{12}(r)$ and $\overline{g}_{21}(r)$, but this effect seems to { be} rather weak. On the contrary, the minority pair increases its correlation function $\overline{g}_{11}(r)$, while the majority pair decreases its correlation function $\overline{g}_{22}(r)$ in the region $1\leq r\lesssim 1.5$ when the composition becomes more balanced. Again, this can be qualitatively understood by the exact results to first order in density [see Eq.\ \eqref{B025}].

\subsection{Preliminary results on the critical behavior}
\label{subsec:phase2}

\begin{figure}[htbp]
\includegraphics[width=8.cm]{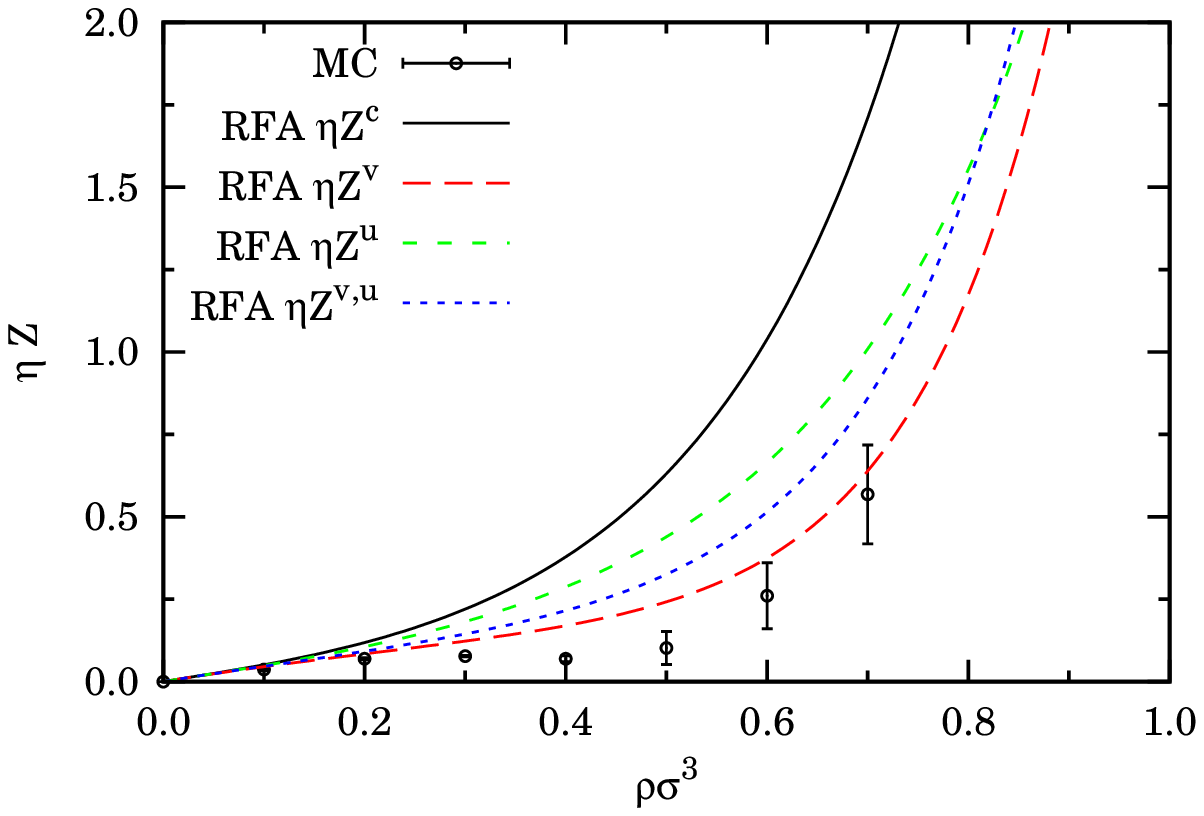}\\
\includegraphics[width=8.cm]{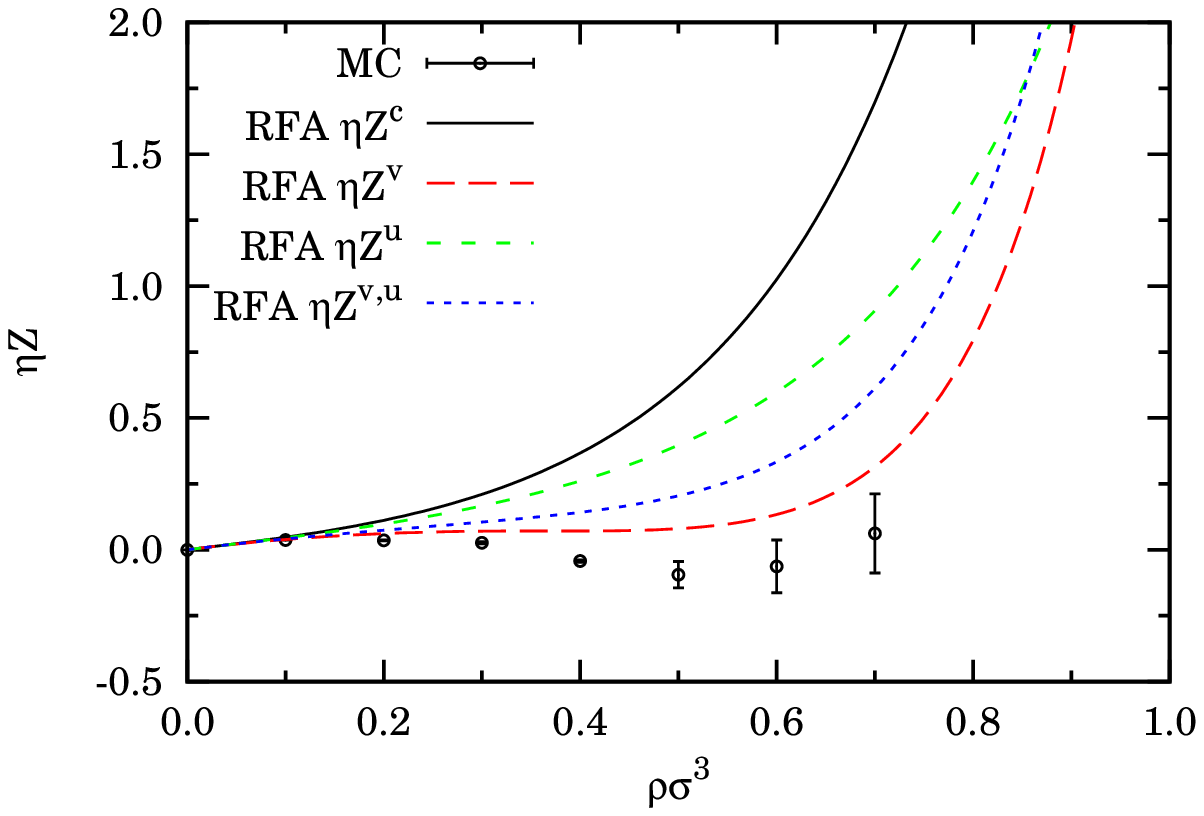}\\
\includegraphics[width=8.cm]{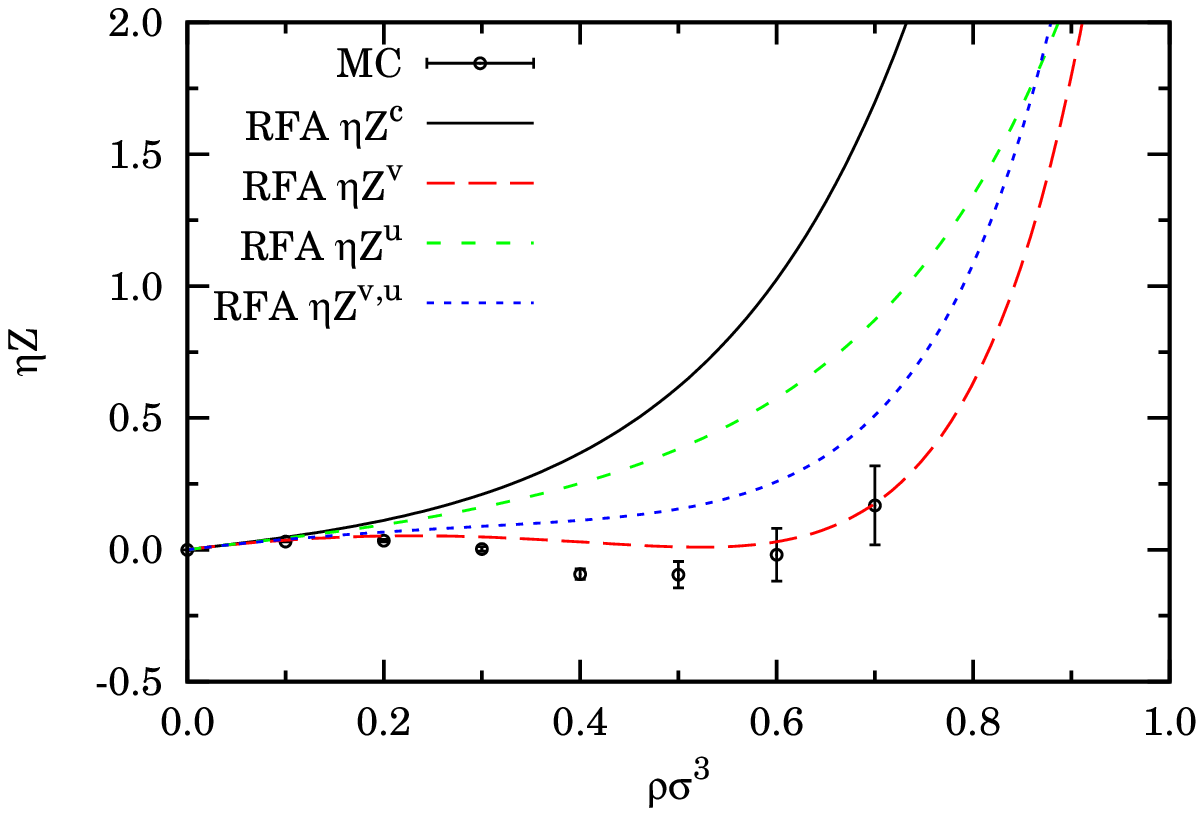}
\caption{MC simulation data for the scaled pressure $\eta Z=\frac{\pi}{6}\sigma^3\beta P$ as a function of $\rho^*$ at $\tau=0.030$ (top panel), $0.0205$ (middle panel), and $\tau=0.018$ (bottom panel) in an equimolar mixture.  Densities higher than $\rho^*=0.7$ are not shown because at these very low temperatures the particles tend to overlap  their SW shells and then the calculations
slow down considerably. Also shown are the theoretical results for the
  four routes of the RFA.}
\label{fig:fig9}
\end{figure}

One of the most interesting and { intriguing} predictions of the RFA is the existence of a {gas}-liquid transition in the equimolar model, despite the fact that only one of the four classes of interactions is attractive (see Sec.\ \ref{subsec:phase}). The elusiveness of this prediction is reflected by the fact that the compressibility route does not account for a critical point and, although the virial and energy routes do, they widely differ in its location, as seen in Table \ref{tab1}. In this region of very low values of $\tau$ the hybrid virial-energy equation of state is dominated
by the virial one and thus the corresponding critical point is not far from the virial { value}.

A simple heuristic argument can be used to { support the existence of a true} critical point in our model. According to the Noro--Frenkel (NF) generalized principle of corresponding states,\cite{NF00} the critical temperatures of different systems of particles interacting via a hard-core potential plus a short-range attraction are such that the reduced second virial coefficient $B_2^*=B_2/B_2^\hs$ has a \emph{common} value $B_2^{*c}\simeq -1.21$. In our model, the reduced second virial coefficient is $B_2^*=1-3t/4=1-1/16\tau$ [see Eq.\  \eqref{B030}]. Thus, assuming {the} NF ansatz, the critical point would correspond to $\tau_c^\text{NF}\simeq 0.028$, a  value higher than but comparable to that listed in Table \ref{tab1} from the virial route.

From the computational point of view, { a direct assessment on the existence of a gas-liquid transition in the present model is not a straightforward task. Unlike the original SHS Baxter model,  a Gibbs ensemble MC (GEMC) calculation for a binary mixture is required to find the coexistence lines. We are currently pursuing this analysis that will be reported elsewhere. As a very preliminary {study}, we here report NVT results  with} values of the Baxter temperature close to the critical value $\tau_c^\text{NF}\simeq 0.028$ expected on the basis of the NF conjecture. More specifically, we have considered $\tau=0.030$, $0.0205$, and $0.018$ (corresponding to $T^*\simeq 0.251$, $0.229$, and $0.223$, respectively). The numerical results for the pressure, along with the RFA theoretical predictions, are displayed in Fig.\ \ref{fig:fig9}.

We observe that at $\tau=0.030$ (top panel) the four theoretical routes clearly indicate a single-phase
gas-like behavior with a monotonic increase of the pressure as a function of the density, in consistence with the value $\tau_c\simeq 0.0205$ obtained from the RFA virial route. On the other hand, the MC data show a practically constant pressure between $\rho^*=0.2$ and $\rho^*=0.4$, which is suggestive of $\tau=0.030$ being close to the critical isotherm (remember that $\tau_c^\text{NF}\simeq 0.028$).
The middle panel has been chosen to represent the critical isotherm predicted by the RFA-virial equation of state. In that case, the simulation data present a clear van der Waals loop with even negative pressures around the minimum. A similar behavior is observed at $\tau=0.018$ (bottom panel), except that now the RFA-virial isotherm also presents a visible van der Waals loop. Whereas the observation of negative values { of isothermal compressibility} in the MC simulations can be attributed to finite-size effects and are expected to disappear in the thermodynamic limit, these preliminary results are highly supportive of the existence of a {gas}-liquid phase transition in our model with a critical (Baxter) temperature $\tau_c\approx 0.03$.

{
In view of the extremely short-range nature of the potential, the stability of the above liquid phases with respect to the corresponding solid ones may be rightfully questioned.\cite{Sciortino10} This is a general issue -- present even in the original Baxter model, as well as
in the spherically symmetric SW or Yukawa potentials with sufficiently small {interaction range}\cite{BHD94,LA94,HF94,MN94} -- that is clearly outside the scope of the present manuscript. In any case, it seems reasonable to expect that at sufficiently low temperature and high density the stable phase will consist of an fcc crystal made of layers  of alternating species (1-2-1-2-$\cdots$) along the $z$ direction.
}

\section{Conclusions and future perspectives}
\label{sec:conclusions}
We have studied thermophysical and structural properties of a Janus fluid
having { constrained} orientations for the attractive hemisphere. The Janus fluid has been modeled using a Kern--Frenkel
potential with a single SW patch pointing either up or down, and studied using numerical  NVT MC simulations.

The above model has been  mapped onto {an} { asymmetric} binary mixture where
the only memory of the original anisotropic potential { stems from} the relative
position along the $z$-axis of particles of the two species $1$ and $2$.
In this way, only one {[$(1,2)$ with our choice of labels]} out of the four possible interactions
is attractive, the {other ones} {[$(1,1)$, $(2,1)$, and $(2,2)$]} being simply HS
interactions.

In the limit of infinitely short and deep SW interactions (sticky limit),
we discussed how a full analytical { theory} is possible.
{ We} developed a new formulation { for asymmetric mixtures} of the rational-function approximation (RFA), that is equivalent to the PY approximation
in the case of symmetric SHS interactions, but differs from it in the { asymmetric case}.
Results from this theory were shown to be in nice agreement with MC simulations
using SW interactions of sufficiently short width ($5\%$ of particle
size), both for the structural and the thermodynamic properties.

The above agreement is rather remarkable in view of the rather severe approximations
involved in the RFA analysis ---that are however largely compensated by the
possibility of a full analytical treatment--- and, more importantly, by the fact
that simulations {deal} with the actual Kern--Frenkel potential with up-down
constrained orientations of the patches and SW attractions, while the RFA theory deals with the obtained
asymmetric binary mixture and SHS interactions.
We regard this agreement as an important indication on the correctness of the mapping.

Within the RFA approximation, all three standard routes to thermodynamics (compressibility, virial,
and energy) were considered. To them we added a weighted average of the virial and energy routes with a weight fixed as to reproduce the exact third virial coefficient. Somewhat surprisingly, our results indicate that only the compressibility
route fails to display a full critical behavior with a well-defined critical point.
The existence of a critical point and a {(possibly metastable)} { gas}-liquid phase transition in our model (despite the fact that attractive interactions are { partially} inhibited) are supported by the NF generalized principle of corresponding states\cite{NF00} and by { preliminary simulations results}. We plan to carry out more detailed GEMC simulations to fully elucidate this point.

The work presented here can { foster further activities} toward an analytical theory of the anomalous phase diagram
indicated by numerical simulations of the (unconstrained) Janus fluid. { We are currently}
working on the extension of the present model allowing for more general interactions {where the red  hemispheres in Fig.\ \ref{fig:fig2}  also present a certain adhesion (e.g., $\tau_{12}<\tau_{11}=\tau_{22}=\tau_{21}<\infty$). This more general model (to which the theory presented in Sec.\ \ref{subsec:rfa} still applies)
can be continuously tuned from isotropic SHS ($\tau_{ij}=\tau$) to isotropic HS interactions ($\tau_{ij}\to\infty$). The increase in
the (Baxter) critical temperatures and densities occurring when equating the stickiness of both hemispheres { would then mimic}
the corresponding increase in the location of the critical point
upon an increase of the patch coverage in the Kern--Frenkel model}.\cite{Sciortino10}

\begin{acknowledgments}
{The research of M.A.G.M. and A.S. has been supported by the Spanish government  through Grant No.\ FIS2010-16587 and by the Junta de Extremadura (Spain) through Grant No.\ GR101583, partially financed by FEDER funds. M.A.G.M is also grateful to the Junta de Extremadura (Spain) for  the pre-doctoral fellowship PD1010.
A.G. acknowledges financial support by PRIN-COFIN 2010-2011 (contract 2010LKE4CC).
R.F. would like to acknowledge the use of the computational facilities of CINECA through the ISCRA call.}
\end{acknowledgments}
\appendix
\section{Exact low-density properties for anisotropic SHS mixtures}
\label{app:appa}
\subsection{Cavity function to first order in density}
To first order in density, the cavity function of an anisotropic mixture is
\beq
y_{ij}(\mathbf{r})=1+y_{ij}^{(1)}(\rr)\rho+\mathcal{O}(\rho^2),
\label{B0}
\eeq
where
\beq
y_{ij}^{(1)}(\rr)=\sum_k x_k y_{ij;k}^\one(\rr),
\label{B1}
\eeq
with
\beq
y_{ij;k}^\one(\rr)=\int d\rr'\, f_{ik}(\rr')f_{kj}(\rr-\rr').
\label{B2}
\eeq
Here,
$
{f_{ij}(\rr)=e^{-\beta \phi_{ij}(\rr)}-1}
$
is the Mayer function.
In the particular case of the anisotropic SHS potential considered in this paper,
\beqa
f_{ij}(\mathbf{r})&=&f_\hs(r)+\delta(r-1)\left[t_{ij}\Theta(\cos\theta)+t_{ji}\Theta(-\cos\theta)\right]\nn
&=&f_{ji}^\shs(r)+t_{ij}^-\delta(r-1)\Theta(\cos\theta),
\label{B3}
\eeqa
where
$
{t_{ij}^-\equiv t_{ij}-t_{ji},}
$
\beq
f_\hs(r)=-\Theta(1-r),\quad f_{ji}^\shs(r)=f_\hs(r)+t_{ji}\delta(r-1).
\label{B4}
\eeq
Inserting Eq.\ \eqref{B3} into Eq.\ \eqref{B2}, we get
\beqa
y_{ij;k}^\one(\rr)&=&\Theta(2-r)\left\{\frac{\pi}{12}(2-r)^2(4+r)
-(t_{ki}+t_{jk})\pi(2-r)\right.\nn
&&+t_{ki}t_{jk}2\pi\left[2\delta(r)+\frac{1}{r}\right]-(t_{ik}^-+t_{kj}^-)\mathcal{A}(\mathbf{r})\nn
&&\left.+(t_{ik}^-t_{jk}+t_{kj}^-t_{ki})\mathcal{L}(\rr)+t_{ik}^-t_{kj}^-\mathcal{L}_0(\rr)\right\},
\label{B5}
\eeqa
where
\beq
\mathcal{A}(\mathbf{r})\equiv \int d\rr'\,\delta(r'-1)\Theta(1-|\rr-\rr'|)\Theta(z'),
\label{B6a}
\eeq
\beq
\mathcal{L}(\mathbf{r})\equiv \int d\rr'\,\delta(r'-1)\delta(|\rr-\rr'|-1)\Theta(z'),
\label{B6b}
\eeq
\beq
\mathcal{L}_0(\mathbf{r})\equiv \int d\rr'\,\delta(r'-1)\delta(|\rr-\rr'|-1)\Theta(z')\Theta(z-z').
\label{B6c}
\eeq

It can be proved that
\beq
\mathcal{A}(\mathbf{r})=\begin{cases}
  \pi(2-r),&\sqrt{1-r^2/4}\leq\cos\theta\leq 1,\\
  A(r/2,\theta),& |\cos\theta|\leq \sqrt{1-r^2/4},\\
  0,& -1\leq \cos\theta\leq -\sqrt{1-r^2/4},
\end{cases}
\label{B7}
\eeq
\beq
\mathcal{L}(\mathbf{r})=\begin{cases}
  {2\pi}/{r},&\sqrt{1-r^2/4}\leq\cos\theta\leq 1,\\
  L(r/2,\theta),& |\cos\theta|\leq \sqrt{1-r^2/4},\\
  0,& -1\leq \cos\theta\leq -\sqrt{1-r^2/4},
\end{cases}
\label{B9}
\eeq
\beq
\mathcal{L}_0(\mathbf{r})=\begin{cases}
  {2\pi}/{r},&\sqrt{1-r^2/4}\leq\cos\theta\leq 1,\\
  L_0(r/2,\theta),& 0\leq\cos\theta\leq \sqrt{1-r^2/4},\\
  0,&  \cos\theta\leq 0,
\end{cases}
\label{B11}
\eeq
where
\beqa
A(\dl,\theta)&=&2\pi\Theta(\cos\theta)-\pi\dl-2\dl\sin^{-1}\frac{\dl\cos\theta}{\sqrt{1-\dl^2}\sin\theta}\nn
&&-2\tan^{-1}\frac{\sqrt{\sin^2\theta-\dl^2}}{\cos\theta},
\label{B8}
\eeqa
\beqa
L(\dl,\theta)&=&-\frac{1}{2\dl}\frac{\partial}{\partial \dl}A(\dl,\theta)\nn
&=&\frac{1}{\dl}\left[\frac{\pi}{2}+\sin^{-1}\frac{\dl\cos\theta}{\sqrt{1-\dl^2}\sin\theta}\right],
\label{B10}
\eeqa
\beqa
L_0(\dl,\theta)&=&L(\dl,\theta)-L(\dl,\pi-\theta)\nn
&=&\frac{2}{\dl}\sin^{-1}\frac{\dl\cos\theta}{\sin\theta\sqrt{1-\dl^2}}.
\label{B12}
\eeqa
In Eqs.\ \eqref{B9} and \eqref{B11} we have omitted terms proportional to $\delta(r)$ since they will not contribute to $g_{ij}(\rr)$.
Note the symmetry relations $\mathcal{A}(\rr)+\mathcal{A}(-\rr)=\pi(2-r)$, $\mathcal{L}(\rr)+\mathcal{L}(-\rr)=2\pi/r$,
$\mathcal{L}(\rr)-\mathcal{L}(-\rr)=\mathcal{L}_0(\rr)-\mathcal{L}_0(-\rr)$.

The orientational average
\beq
\overline{y}_{ij;k}^\one(r)=\int_0^{\pi/2} d\theta\,\sin\theta y_{ij;k}^\one(\rr)
\label{B13}
\eeq
becomes
\beqa
\overline{y}_{ij;k}^\one(r)&=&\Theta(2-r)\left\{\frac{\pi}{12}(2-r)^2(4+r)-(t_{ki}+t_{jk})\pi(2-r)\right.\nn
&&+t_{ki}t_{jk}2\pi\left[2\delta(r)+\frac{1}{r}\right]-(t_{ik}^-+t_{kj}^-)\overline{\mathcal{A}}({r})\nn
&&\left.+(t_{ik}^-t_{jk}+t_{kj}^-t_{ki})\overline{\mathcal{L}}(r)+t_{ik}^-t_{kj}^-\overline{\mathcal{L}}_0(r)\right\},
\label{B14}
\eeqa
where
\beq
\overline{\mathcal{A}}({r})=\pi(2-r)\left(1-\sqrt{1-r^2/4}\right)+\overline{A}(r/2),
\label{B15}
\eeq
\beq
\overline{\mathcal{L}}({r})=\frac{2\pi}{r}\left(1-\sqrt{1-r^2/4}\right)+\overline{L}(r/2),
\label{B16}
\eeq
\beq
\overline{\mathcal{L}}_0({r})=\frac{2\pi}{r}\left(1-\sqrt{1-r^2/4}\right)+\overline{L}_0(r/2),
\label{B17}
\eeq
with
\beqa
\overline{A}(\dl)&=&\int_{\sin^{-1}\dl}^{\pi/2}d\theta\,\sin\theta A(\dl,\theta)\nn
&=&2\sqrt{1-\dl^2}\left(\pi-\pi\dl-1\right)+2\dl\cos^{-1}\dl,
\label{B18}
\eeqa
\beqa
\overline{L}(\dl)&=&\int_{\sin^{-1}\dl}^{\pi/2}d\theta\,\sin\theta L(\dl,\theta)\nn
&=&\frac{1}{\dl}\left({\pi}\sqrt{1-\dl^2}-\cos^{-1}\dl\right),
\label{B19}
\eeqa
\beqa
\overline{L}_0(\dl)&=&\int_{\sin^{-1}\dl}^{\pi/2}d\theta\,\sin\theta L_0(\dl,\theta)\nn
&=&\frac{1}{\dl}\left({\pi}\sqrt{1-\dl^2}-2\cos^{-1}\dl\right).
\label{B20}
\eeqa

\subsection{Second and third virial coefficients}
\label{app:appb}
The series expansion of the compressibility factor $Z$ in powers of density defines the virial coefficients:
\beq
Z=1+B_2\rho+B_3\rho^2+\cdots.
\label{B23}
\eeq
Using Eq.\ \eqref{B0} in Eq.\ \eqref{analytical:eq6}, one can identify
\beq
B_2=\frac{2\pi}{3}\left(1-3\sum_{i,j}x_i x_j t_{ij}\right),
\label{B24}
\eeq
\beq
B_3=\frac{2\pi}{3}\sum_{i,j,k}x_i x_j x_k\left[(1-3t_{ij})\overline{y}_{ij;k}^\one(1)-t_{ij}{\overline{y}_{ij;k}^\one}'(1)\right].
\label{B25}
\eeq
According to Eq.\ \eqref{B14},
\beqa
\overline{y}_{ij;k}^\one(1)&=&\frac{5\pi}{12}-(t_{ki}+t_{jk})\pi+t_{ki}t_{jk}2\pi
-(t_{ik}^-+t_{kj}^-)\overline{\mathcal{A}}({1})\nn
&&+(t_{ik}^-t_{jk}+t_{kj}^-t_{ki})\overline{\mathcal{L}}(1)+t_{ik}^-t_{kj}^-
\overline{\mathcal{L}}_0(1),
\label{B26}
\eeqa
\beqa
{\overline{y}_{ij;k}^\one}'(1)&=&-\frac{3}{4}\pi+(t_{ki}+t_{jk})\pi-t_{ki}t_{jk}2\pi-(t_{ik}^-+t_{kj}^-)\overline{\mathcal{A}}'({1})
\nn
&&+
(t_{ik}^-t_{jk}+t_{kj}^-t_{ki})\overline{\mathcal{L}}'(1)+t_{ik}^-t_{kj}^-
\overline{\mathcal{L}}_0'(1),
\label{B27}
\eeqa
where
\beq
\overline{\mathcal{A}}(1)=\frac{4\pi}{3}-\sqrt{3},\quad
\overline{\mathcal{A}}'(1)=-\frac{2\pi}{3},
\label{B21}
\eeq
\beq
\overline{\mathcal{L}}(1)=\frac{4\pi}{3},\quad \overline{\mathcal{L}}'(1)=-\frac{2}{3}\left(2\pi-\sqrt{3}\right),
\label{B22a}
\eeq
\beq
\overline{\mathcal{L}}_0(1)=\frac{2\pi}{3},
\quad \overline{\mathcal{L}}_0'(1)=-\frac{2}{3}\left(\pi-2\sqrt{3}\right).
\label{B22}
\eeq

The second and third virial coefficients can also { be} obtained from the compressibility equation \eqref{orientational:eq13}. To that end, note that
\beq
\widehat{h}_{ij}(0)=\widehat{h}_{ij}^\one(0)\rho+\widehat{h}_{ij}^\two(0)\rho^2+\cdots,
\label{hij0}
\eeq
where, according to Eq.\ \eqref{orientational:eq12},
\beq
\widehat{h}_{ij}^\one(0)=\sqrt{x_i x_j}2\pi\left(-\frac{2}{3}+t_{ij}+t_{ji}\right),
\label{hii01}
\eeq
\beqa
\widehat{h}_{ij}^\two(0)&=&\sqrt{x_i x_j}2\pi\Big\{t_{ij}\overline{y}_{ij}^\one(1)+t_{ji}\overline{y}_{ji}^\one(1)\nn
&&+\int_1^2dr\, r^2\left[\overline{y}_{ij}^\one(r)+\overline{y}_{ji}^\one(r)\right]\Big\}.
\label{hii02}
\eeqa
Inserting this into Eq.\ \eqref{orientational:eq13} and making use of Eqs.\ \eqref{B14}--\eqref{B20}, one gets $\chi_T^{-1}=1+2B_2\rho+3B_3\rho^2+\cdots$, with $B_2$ and $B_3$ given by Eqs.\ \eqref{B24} and \eqref{B25}, respectively.
Furthermore, it can be checked that the exact consistency condition \eqref{consistency} is satisfied by Eqs.\ \eqref{B0}, \eqref{B1}, \eqref{B26}, and \eqref{B27}. The verification  of these two thermodynamic consistency conditions  represent stringent tests on the correctness of the results derived in this appendix.

\subsection{Case $t_{11}=t_{22}=t_{21}=0$}
In the preceding equations of this appendix we have assumed general values for the stickiness parameters $t_{ij}$. On the other hand,  significant simplifications occur in our constrained Janus model, where $t_{ij}=t\delta_{i1}\delta_{j2}$. More specifically,
\beqa
y_{11}^\one(\rr)&=&\Theta(2-r)\Big\{\frac{\pi}{12}(2-r)^2(4+r)-x_2 t\left[\pi(2-r)\right.\nn
&&\left.-t\mathcal{L}(\rr)+t\mathcal{L}_0(\rr)\right]\Big\},
\label{B023}
\eeqa
\beq
y_{12}^\one(\rr)=\Theta(2-r)\left[\frac{\pi}{12}(2-r)^2(4+r)-t
\mathcal{A}(\rr)\right],
\label{B024}
\eeq
\beqa
\overline{y}_{11}^\one(r)&=&\Theta(2-r)\Big[\frac{\pi}{12}(2-r)^2(4+r)-x_2 \pi t\Big(2-r\nn
&&-\frac{2t}{\pi r}\cos^{-1}\frac{r}{2}\Big)\Big],
\label{B025}
\eeqa
\beqa
\overline{y}_{12}^\one(r)&=&\Theta(2-r)\left\{\frac{\pi}{12}(2-r)^2(4+r)-t\Big[\pi(2-r)\right.\nn
&&\left.-2\sqrt{1-r^2/4}+r\cos^{-1}
\frac{r}{2}\Big]\right\},
\label{B026}
\eeqa
\beqa
\overline{y}_{21}^\one(r)&=&\Theta(2-r)\Big\{\frac{\pi}{12}(2-r)^2(4+r)-t\Big[2\sqrt{1-r^2/4}\nn
&&-r\cos^{-1}
\frac{r}{2}\Big]\Big\},
\label{B027}
\eeqa
\beqa
\overline{y}^\one(r)&=&\Theta(2-r)\Big[\frac{\pi}{12}(2-r)^2(4+r)-x_1x_2 2\pi t\Big(2-r\nn
&&-\frac{t}{\pi r}\cos^{-1}\frac{r}{2}\Big)\Big],
\label{B025b}
\eeqa
\beq
\overline{y}_{11}^\one(1)=\frac{5\pi}{12}-x_2\pi t\left(1-\frac{2t}{3}\right),
\eeq
\beq
\overline{y}_{12}^\one(1)=\frac{5\pi}{12}- t\left(\frac{4\pi}{3}-\sqrt{3}\right),
\eeq
\beq
\overline{y}_{21}^\one(1)=\frac{5\pi}{12}- t\left(\sqrt{3}-\frac{\pi}{3}\right),
\label{B028}
\eeq
\beq
{\overline{y}_{11}^\one}'(1)=-\frac{3\pi}{4}+x_2 t\left[\pi-\frac{2t}{3}\left(\pi+\sqrt{3}\right)\right],
\eeq
\beq
{\overline{y}_{12}^\one}'(1)=-\frac{3\pi}{4}+ t\frac{2\pi}{3},
\eeq
\beq
{\overline{y}_{21}^\one}'(1)=-\frac{3\pi}{4}+ t\frac{\pi}{3},
\label{B029}
\eeq
\beq
\overline{y}^\one(1)=\frac{5\pi}{12}-x_1x_22\pi t\left(1-\frac{t}{3}\right),
\eeq
\beq
{\overline{y}^\one}'(1)=-\frac{3\pi}{4}+x_1x_2 2t\left[\pi-\frac{t}{3}\left(\pi+\sqrt{3}\right)\right],
\label{B028b}
\eeq
\beq
\frac{6}{\pi}B_2=4\left(1-3x_1x_2 t\right),
\label{B030}
\eeq
\beq
\left(\frac{6}{\pi}\right)^2B_3=10\left\{1-6x_1 x_2 t\left[1-\frac{2}{5}\left(4-3\frac{\sqrt{3}}{\pi}\right)t\right]\right\},
\label{B031}
\eeq
\beqa
\frac{u_\ex}{\epsilon}&=&-12\eta x_1 x_2 t\left\{1+\frac{5}{2}\left[1-\frac{4}{5}\left(4-3\frac{\sqrt{3}}{\pi}\right)t\right]\eta\right\}
\nn&&+\mathcal{O}(\eta^2).
\label{B032}
\eeqa

\section{Evaluation of the coefficients $L_{ij}^\zero$, $L_{ij}^\one$, and $L_{ij}^\two$}
\label{appB}
In order to apply Eqs.\ \eqref{III.1a} and \eqref{III.1b}, it is convenient to rewrite Eq.\ \eqref{analytical:eq19}  as
\begin{eqnarray}
\label{analytical:eq23}
\frac{1}{2\pi}\mathsf{L}(s)&=&\mathsf{Q}(s)\cdot \left[\mathsf{I}-
\mathsf{A}(s)\right],
\end{eqnarray}
where we have introduced the matrix $\mathsf{Q}$ as
\begin{eqnarray}
\label{analytical:eq24}
Q_{ij}(s)&\equiv& e^{s}s^2 G_{ij}(s).
\end{eqnarray}
Thus, Eqs.\ \eqref{III.1a} and \eqref{III.1b} are equivalent to
\begin{eqnarray}
\label{analytical:eq25}
Q_{ij}(s)&=&1+s+\mathcal{O}(s^2).
\end{eqnarray}
Expanding $A_{ij}(s)$ in powers of $s$ and inserting {the result} into Eq.\ \eqref{analytical:eq23}, one gets
\begin{eqnarray}
\label{analytical:eq28}
\frac{1}{2\pi}L_{ij}^\zero&=&1-\sum_{k} A_{kj}^\zero,
\end{eqnarray}
\begin{eqnarray}
\label{analytical:eq29}
\frac{1}{2\pi}L_{ij}^\one&=&1-\sum_{k} \left(A_{kj}^\one
+A_{kj}^\zero\right),
\end{eqnarray}
where
\beq
\label{analytical:eq27}
 A_{ij}^\n =(-1)^n\rho x_i \left[\frac{L_{ij}^\zero}{(n+3)!}
 -\frac{L_{ij}^\one}{(n+2)!}+
 \frac{L_{ij}^\two}{(n+1)!}\right].
 \eeq
Equations (\ref{analytical:eq28}) and (\ref{analytical:eq29}) constitute a {\em linear} set of equations which allow us to express the coefficients $L_{ij}^\zero$ and $L_{ij}^\one$ in terms of the coefficients {$\{L_{kj}^\two\}$}. The result is given by Eqs.\ \eqref{analytical:eq30} and \eqref{analytical:eq31}.

It now remains the determination of $L_{ij}^\two$. This is done by application of Eq.\ \eqref{IV.1}, i.e., the ratio
first term to second term in the expansion of $e^sG_{ij}(s)$
for large $s$ must be exactly equal to $t_{ij}$. This is the only point where the stickiness parameters of the mixture appear explicitly.

The large-$s$ behavior from Eq.\ (\ref{analytical:eq19}) is
\beq
\label{analytical:eq32}
2\pi e^{s}G_{ij}(s)=L_{ij}^\two+\left[
L_{ij}^\one+\left({\sf L}^\two\cdot {\sf D}\right)_{ij}\right]s^{-1}
+{\cal O}(s^{-2}),
\eeq
where
\begin{eqnarray}
\label{analytical:eq33}
D_{ij}&\equiv& \rho x_i\left(\frac{1}{2} L_{ij}^\zero-
 L_{ij}^\one+L_{ij}^\two\right)\nn
&=&\rho x_i\left(L_{ij}^\two-\frac{\pi}{1-\eta}\right).
\end{eqnarray}
Comparison of Eq.\ (\ref{analytical:eq13}) with Eq.\ (\ref{analytical:eq32}) yields Eq.\ \eqref{analytical:eq34} and
\begin{eqnarray}
\label{analytical:eq35a}
\frac{12\tau_{ij}L_{ij}^\two}{\sigma_{ij}}&=&
L_{ij}^\one+\sum_{k=1}^m L_{ik}^\two D_{kj}.
\end{eqnarray}
\begin{eqnarray}
\label{analytical:eq35}
\frac{L_{ij}^\two}{t_{ij}}&=&
L_{ij}^\one+\sum_{k} L_{ik}^\two D_{kj}.
\end{eqnarray}
Taking into account Eqs.\ (\ref{analytical:eq31}) and (\ref{analytical:eq33}), Eq.\ (\ref{analytical:eq35}) becomes Eq.\ \eqref{analytical:eq36}.

\section{{Recovery of the pseudo-PY solution}}
\label{appC}
The aim of this appendix is to prove that the pair correlation functions $\overline{g}_{ij}(r)$ obtained from the RFA method in Sec.\ \ref{subsec:rfa} satisfy Eqs.\ \eqref{analytical:eq16} and \eqref{analytical:eq18}.

First, note that the pseudo-OZ relation \eqref{analytical:eq18} can be rewritten in the form
\beq
\widehat{\overline{\mathsf{c}}}(q)=\widehat{\overline{\mathsf{h}}}(q)\cdot\left[\mathsf{I}+\widehat{\overline{\mathsf{h}}}(q)\right]^{-1},
\label{OZ}
\eeq
where $\mathsf{I}$ is the  unit matrix and
\beq
\widehat{\overline{c}}_{ij}(q)=\rho \sqrt{x_i x_j}\int d\rr\, e^{-i\mathbf{q}\cdot\rr}\overline{c}_{ij}(r),
\label{C1a}
\eeq
\beq
\widehat{\overline{h}}_{ij}(q)=\rho \sqrt{x_i x_j}\int d\rr\, e^{-i\mathbf{q}\cdot\rr}\overline{h}_{ij}(r).
\label{C1b}
\eeq
Note that $\widehat{h}_{ij}(0)=\frac{1}{2}\left[\widehat{\overline{h}}_{ij}(0)+\widehat{\overline{h}}_{ji}(0)\right]$, where $\widehat{h}_{ij}(0)$ is defined by Eq.\ \eqref{orientational:eq12}.

The Fourier transform $\widehat{\overline{h}}_{ij}(q)$ of the (orientational average) total correlation function $\overline{h}_{ij}(r)=\overline{g}_{ij}(r)-1$ is related to the Laplace transform $G_{ij}(s)$ [see Eq.\ \eqref{orientational:eq14}] by
\beq
\label{hij}
\widehat{\overline{h}}_{ij}(q)= -2\pi\rho\sqrt{x_i x_j}
\left[ \frac{G_{ij}(s)-G_{ij}(-s)}{s}\right]_{s=i q}.
\eeq
Making use of Eqs.\  \eqref{analytical:eq19}--\eqref{analytical:eq21}, it is possible to obtain, after some algebra,
\beqa
\frac{\widehat{\overline{c}}_{ij}(q)}{\rho\sqrt{x_ix_j}}&=&\frac{4\pi}{q^3}C_{ij}^\zero\left(\sin q-q\cos q\right)+\frac{4\pi}{q^4}C_{ij}^\one\left[2q\sin q\right.\nn
&&\left.-2-\left(q^2-2\right)\cos q\right]+\frac{4\pi}{q^6}C_{ij}^\three\left[4q\left(q^2-6\right)\right.\nn
&&\left.\times \sin q+24-\left(24-12q^2+q^4\right)\cos q\right]\nn
&&+4\pi t_{ij}{y_{ij}(1)}\frac{\sin q}{q},
\label{3.35}
\eeqa
where the coefficients $C_{ij}^\zero$, $C_{ij}^\one$,  and $C_{ij}^\three$ are independent of $q$ but depend on the density, the composition, and the stickiness parameters. Fourier inversion yields
\beqa
\overline{c}_{ij}(r)&=&\left[C_{ij}^\zero+C_{ij}^\one r+C_{ij}^\three r^3\right]\Theta (1-r)\nn
&&+{y_{ij}(1)}t_{ij}\delta_+(r-1).
\label{3.36}
\eeqa
Taking into account Eq.\ \eqref{analytical:eq8} we see that Eq.\ \eqref{3.36} has the structure
\beq
\overline{c}_{ij}(r)=\overline{g}_{ij}(r)-\overline{y}_{ij}(r).
\label{3.37}
\eeq
But this is not but the PY closure relation \eqref{analytical:eq16}. In passing, we get the cavity function inside the core:
\beq
y_{ij}(r)\Theta(1-r)=-\left[C_{ij}^\zero+C_{ij}^\one r+C_{ij}^\three r^3\right]\Theta (1-r).
\label{3.38}
\eeq


\bibliographystyle{apsrev}

\end{document}